\documentclass[twocolumn]{article}
\usepackage[utf8]{inputenc}
\usepackage[bookmarks=false]{hyperref}
\usepackage{graphicx}
\usepackage{xcolor}
\usepackage{siunitx} 
\usepackage{amssymb}
\usepackage{amsmath}
\usepackage[normalem]{ulem}
\usepackage[]{lipsum}

\usepackage[a4paper,width=185mm]{geometry}
\usepackage{tabularx}

\begin{document}

\title{Parity-conserving Cooper-pair transport and ideal superconducting diode in planar Germanium}

\author{Marco Valentini$^{1,*}$, Oliver Sagi$^{1}$, Levon Baghumyan$^{1}$, Thijs de Gijsel$^{1,2}$, Jason Jung$^{2}$, Stefano Calcaterra$^{3}$, \\Andrea Ballabio$^{3}$, Juan Aguilera Servin$^{1}$, Kushagra Aggarwal$^{1,4}$,  Marian Janik$^{1}$, Thomas Adletzberger$^{1}$,\\  Rubén Seoane Souto$^{5,6}$, Martin Leijnse$^7$, Jeroen Danon$^{8}$, Constantin Schrade$^{5}$, Erik Bakkers$^{2}$,\\ Daniel Chrastina$^{3}$, Giovanni Isella$^{3}$, Georgios Katsaros$^{1,*}$}

\date{%
    $^1$Institute of Science and Technology Austria, Am Campus 1, 3400 Klosterneuburg, Austria.\\%
    $^2$ Department of Applied Physics, Eindhoven University of Technology, Eindhoven, The Netherlands.\\%
    $^3$ L-NESS, Physics Department, Politecnico di Milano, via Anzani 42, 22100, Como, Italy.\\%
    $^4$ Department of Materials, University of Oxford, Oxford OX1 3PH, United Kingdom.\\%
    $^5$ Center for Quantum Devices, Niels Bohr Institute, University of Copenhagen, 2100 Copenhagen, Denmark.\\%
    $^6$ Instituto de Ciencia de Materiales de Madrid, Consejo Superior de Investigaciones Científicas (ICMM-CSIC), Madrid, Spain.\\%
    $^7$ NanoLund and Solid State Physics, Lund University, Box 118, 22100 Lund, Sweden.\\%
    $^8$ Department of Physics, Norwegian University of Science and Technology, NO-7491 Trondheim, Norway.\\%
     $^*$ Corresponding authors: marco.valentini@ist.ac.at and georgios.katsaros@ist.ac.at .\\[2ex]%
        \today
}

\maketitle
\section*{Abstract}
\textbf{Superconductor/semiconductor hybrid devices have attracted increasing interest in the past years. Superconducting electronics aims to complement semiconductor technology, while hybrid architectures are at the forefront of new ideas such as topological superconductivity and protected qubits. In this work, we engineer the induced superconductivity in two-dimensional germanium hole gas by varying the distance between the quantum well and the aluminum. We demonstrate a hard superconducting gap and realize an electrically and flux tunable superconducting diode using a superconducting quantum interference device (SQUID). This allows to tune the current phase relation (CPR), to a regime where single Cooper pair tunneling is suppressed, creating a $\boldsymbol{ \sin \left( 2 \varphi \right)}$ CPR. Shapiro experiments complement this interpretation and the microwave drive allows to create a diode with $\boldsymbol{ \approx 100 \%}$ efficiency. The reported results open up the path towards integration of spin qubit devices, microwave resonators and (protected) superconducting qubits on a silicon technology compatible platform.}

\begin{figure*} [h!]
  \includegraphics[]{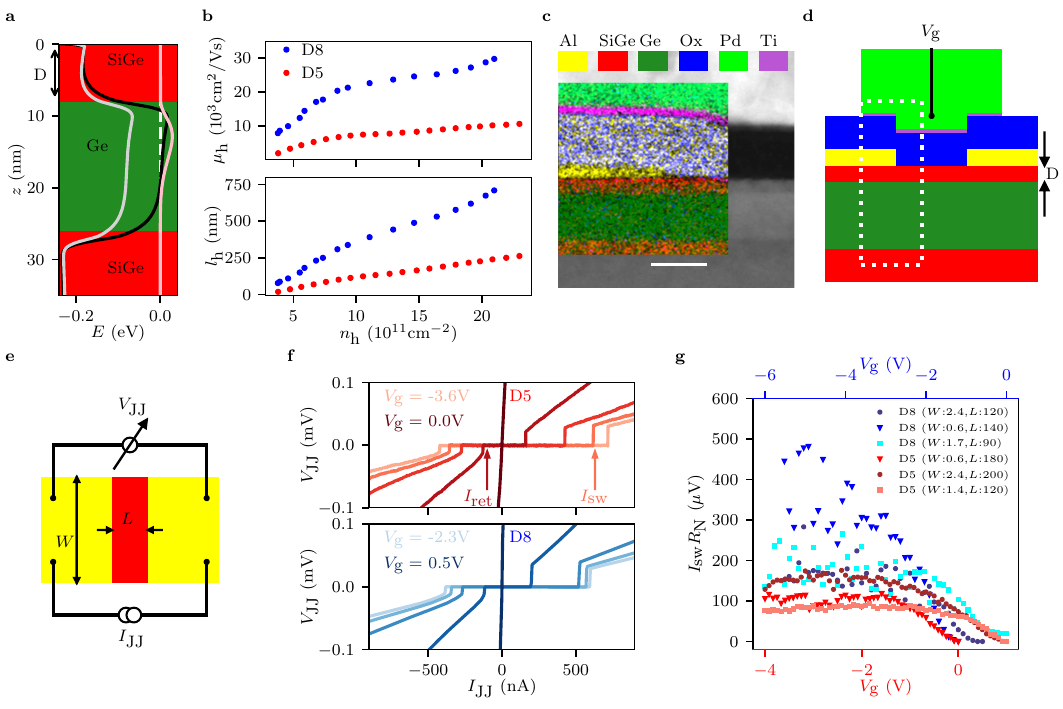}
  \caption{\textbf{Proximity induced superconductivity in Planar Ge}. \small{\textbf{a} Heavy hole (HH) [light hole (LH)] band energies (black trace) [(grey trace)] along the growth direction $z$ simulated using NextNano. HHs are accumulated at the upper QW interface, as shown by the pink trace representing the HH wavefunction density plotted in arbitrary units.   \textbf{b} Hole mobility $\mu_{\textrm{h}}$ [upper panel, extracted with Hall bar measurements)] and  mean free path $l_{\textrm{h}}$ [lower panel], extracted from Hall bar measurements, as a function of carrier density $n_{\textrm{h}}$ for samples with ${\textrm{Si}}_{0.3}{\textrm{Ge}}_{0.7}$ spacer thickness (D) of 5nm (D5) and 8nm (D8).  \textbf{c} TEM image of the upper part of the Ge/SiGe heterostructure. The left inset shows EDX data which confirm the absence of Al both in the spacer and in the Ge QW. The scale bar corresponds to 20 nm. \textbf{d} Cross-section sketch of a JoFET device; the dashed rectangle corresponds to the inset in \textbf{c}. The gate voltage $V_{\textrm{g}}$ is used for varying the hole carrier density in the underlying Ge QW. \textbf{e} Top-view sketch of a JoFET device with the circuit for the 4-probe measurement. The width ($W$) and the channel length ($L$) are indicated.  \textbf{f} Voltage drop $V_{\textrm{JJ}}$ measured as a function of the applied current $I_{\textrm{JJ}}$ for D5 [upper panel] and for D8 [lower panel]. Lighter colors indicate lower values of $V_{\textrm{g}}$ [higher carrier density] and darker colors indicate higher values of $V_{\textrm{g}}$ [lower carrier density]. Traces are equally spaced for both panels. \textbf{g}  $I_{\textrm{sw}} R_{\textrm{N}}$ product as a function of $V_{\textrm{g}}$ for D5 and D8 and for Josephson junctions with different dimensions as indicated in the inset. $W$ is reported in units of $\unit{\mu m}$, while $L$ is in units of $\unit{nm}$.}}  \label{figure1}
\end{figure*}

\section*{Main}
III-V semiconductors have become the materials of choice for realizing high-quality hybrid devices, due to the possibility of growing epitaxial Al on top of them \cite{krogstrup2015epitaxy}. Gate-tunable superconducting and Andreev spin qubits \cite{larsen2015semiconductor,de2015realization,casparis2018superconducting,hays2021asq,pita2022direct}, parametric amplifiers \cite{phan2022semiconductor}, highly efficient Cooper pair splitters \cite{wang2022singlet,bordoloi2022spin,wang2022triplet} and a minimal Kitaev chain \cite{dvir2022mini} are prominent examples of what has been achieved in the past decade. In addition, non-reciprocal devices, such as superconducting diodes have attracted a lot of interest \cite{ando2020observation}, especially in Josephson junctions in the presence \cite{baumgartner2022supercurrent,pal2022josephson,mazur2022gate} or absence \cite{steiner2023diode,wu2022field,trahms2023diode} of a Zeeman field  and in multiterminal devices \cite{gupta2022superconducting,chilestriode}. Diodes can be also realized in a superconducting quantum interference device (SQUID) geometry by exploiting a magnetic flux to achieve time-reversal breaking  \cite{souto2022josephson,fominov2022asymmetric,ciaccia2023gate}. Such SQUIDs can be also used as a building block  to create a protected superconducting qubit by engineering a $\sin \left(2 \varphi \right)$ current phase relation (CPR) \cite{smith2020superconducting,brooks2013protected,gyenis2021experimental,larsen2020parity,schrade2022protected,Maiani}.

One drawback of III-V materials is their non-zero nuclear spin, which, through hyperfine interaction, drastically reduces the electron spin coherence time, limiting therefore the use of hybrid devices in combination with the spin degree of freedom \cite{hays2021asq,pita2022direct}. Germanium, on the other hand, is a material which allows proximity induced superconductivity and has shown great potential for spin qubit devices~\cite{scappucci2021}. Induced superconductivity in germanium was first demonstrated in 0D and 1D systems \cite{xiang2006ge,katsaros2010hybrid}. A few years later, superconductivity was also induced in a two-dimensional Ge hole gas \cite{hendrickx2018,vigneau2019germanium}. Recent works demonstrated how induced superconductivity can be improved in planar germanium, either by using a double superconducting stack \cite{aggarwal2021enhancement} or by annealing platinum contacts \cite{tosato2022hard}. 
Here, using a shallow quantum well (QW) we establish Ge/SiGe heterostructures as an alternative platform to III-V materials for hybrid devices and microwave experiments, opening therefore the path to the coexistence of semiconductor and superconducting qubits.

\begin{figure*}[h!]
  \includegraphics[]{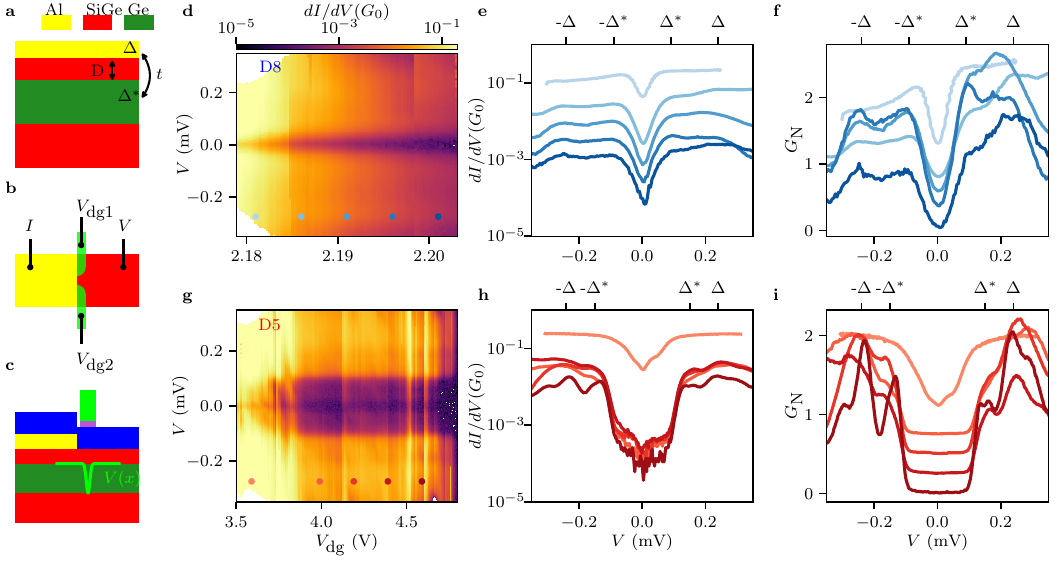}  \caption{\textbf{Superconducting gap tunability.} \small{ \textbf{a} Sketch of the proximity effect. Al has a superconducting parent gap $\Delta$ and it is coupled to the Ge hole gas. The coupling $t$, and therefore the induced gap $\Delta^*$, depends on the thickness of the SiGe tunnel barrier, i.e. on D. \textbf{b} Top-view sketch of the device layout used to perform tunneling spectroscopy. The part of the Ge QW (right side) not covered by Al is tuned to be fully conductive and behaves like a normal metal reservoir. The two split gates are used for creating a tunnel barrier by applying voltages $V_{\textrm{dg1}}$ and $V_{\textrm{dg2}}$. The accumulation gate which covers the sample without Al on top is not depicted in the sketch. \textbf{c} Side-view sketch of \textbf{b}. The green  profile is a sketch of the tunnel barrier for holes formed at the border between the conductive Ge and the hybridized Ge. \textbf{d} [\textbf{g}] $dI/dV$ as a function of $V$ and $V_{\textrm{dg}} = V_{\textrm{dg1}} + V_{\textrm{dg2}}$ plotted in logarithmic scale for D8 [D5]. Data for lower $V_{\textrm{dg}}$ are shown in Fig.~\ref{figure2_SI}. \textbf{e} [\textbf{h}] Line-cuts taken from \textbf{d} [\textbf{g}] at different $V_{\textrm{dg}}$ (see small solid circles) demonstrating a hard gap for sample D5. \textbf{f} [\textbf{i}] Line-cuts taken from \textbf{d} [\textbf{g}] plotted in a normalized scale, in which the measured $dI/dV$ is divided by the normal state conductance $G_{\textrm{N}} = $(dI/dV)$/G_{\textrm{normal}}$. The traces are shifted vertically by 0.25 $G_{\textrm{N}}$ with respect to each other.}}
  \label{figure2}
\end{figure*}

\section*{Material characterization and Josephson junctions}

Compressively strained Ge QWs have been deposited on  relaxed, linearly graded buffers with 70\% Ge content. The 18 nm thick QWs are separated from the top surface by a spacer of thickness D. The built-in in-plane compressive strain leads to charge confinement in the heavy-hole band (see Fig.~\ref{figure1}\textbf{a}). Mobility ($\mu_{\textrm{h}}$) and  mean free path ($l_{\textrm{h}}$) as a function of the carrier density $n_{\textrm{h}}$ are displayed in Fig.~\ref{figure1}\textbf{b} for a QW with $\approx \qty{5}{nm}$ ${\textrm{Si}}_{0.3}{\textrm{Ge}}_{0.7}$ spacer (D5 - red) and $\approx \qty{8}{nm}$ (D8 - blue), respectively. At high density, D5 (D8) shows $\mu_{\textrm{h}} \approx \qty{10000}{ {cm}^2 / V s}$ ($\qty{30000}{ {cm}^2 / V s}$) and $l_{\textrm{h}} \approx \qty{250}{nm}$ ($ \qty{700}{nm}$). 

For creating hybrid superconductor-semiconductor devices a thin film of aluminum ($\approx 8-10 \ \unit{nm}$) is deposited \textit{ex situ} and at low temperature on top of the ${\textrm{Si}}_{0.3}{\textrm{Ge}}_{0.7}$ spacer (see Methods for the growth, the aluminum deposition and fabrication details). The Al has a polymorphic structure and it is not grown epitaxially on top of the  ${\textrm{Si}}_{0.3}{\textrm{Ge}}_{0.7}$ spacer. Importantly, energy dispersive X-ray (EDX) data do not reveal interdiffusion of Al inside the ${\textrm{Si}}_{0.3}{\textrm{Ge}}_{0.7}$ spacer and Ge QW, see Fig.~\ref{figure1}\textbf{c}.

In order to check if the superconducting properties can leak into the Ge hole gas, Josephson field effect transistors (JoFETs) were fabricated  (Fig.~\ref{figure1}\textbf{d}). Representative $V_{\textrm{JJ}}$ vs $I_{\textrm{JJ}}$ traces, measured in a four-terminal current-biased configuration (Fig.~\ref{figure1}\textbf{e}), for D5 [red, upper plot] and D8 [blue, lower plot] are shown in Fig.~\ref{figure1}\textbf{f}. The devices switch from superconducting to the dissipative regime at the gate tunable switching current $I_{\textrm{sw}}$.
A common figure of merit used for  estimating the quality of the proximity effect is the product between $I_{\textrm{sw}}$ and the normal state resistance $R_{\textrm{N}}$. Fig.~\ref{figure1}\textbf{g} reports this product as a function of the gate voltage $V_{\textrm{g}}$ for 6 different junctions with different dimensions.
At high negative values of gate voltages, $I_{\textrm{sw}} R_{\textrm{N}}$ spans from slightly below $\qty{100}{\mu eV}$ to above $\qty{400}{\mu eV}$ depending on D and on the JoFET dimensions.  These values are favourably compared to previous results obtained with Ge heterostructures hybridized by Al \cite{hendrickx2018,vigneau2019germanium} and they are on par with more mature material systems \cite{Mayer2019}.

In sample D5, the proximity effect is expected to be more effective because the Al is closer to the Ge hole gas. It is then surprising that the measured $I_{\textrm{c}} R_{\textrm{N}}$ product shown in Fig.~\ref{figure1}\textbf{g} is significantly smaller for sample D5. One factor that could play a role is the fact that D8, especially at high density, is in the short ballistic regime ($L < l_{\textrm{h}}, \xi_{\textrm{N}}$), where $I_{\textrm{c}} R_{\textrm{N}}$ is expected to be equal to $\pi \Delta / e$ \cite{likharev1979superconducting}; $I_{\textrm{c}}$ is the critical current and $\xi_{\textrm{N}} = \frac{\hbar^2 \sqrt{ 2 \pi n_{\textrm{h}}}}{2 m_{\textrm{eff}} \Delta}$ is the superconducting coherence length in the quantum well with $m_{\textrm{eff}}$ being the effective mass. Using $n_{\textrm{h}} =10^{12} \unit{cm^{-2}}$, $\Delta = \qty{200}{\mu eV}$ and $m_{\textrm{eff}}$ to be around $10 \%$ of the electron mass~\cite{rossner2003effective}, we estimated $\xi_{\textrm{N}} \approx 500 \ \si{n m}$. On the contrary, samples D5 have $L \approx l_{\textrm{h}}$, which implies a smaller $I_{\textrm{c}} R_{\textrm{N}}$ \cite{likharev1979superconducting}, making it challenging to compare the $I_{\textrm{c}} R_{\textrm{N}}$ of the D8 and D5 devices directly. Moreover, the variations of $I_{\textrm{c}} R_{\textrm{N}}$ in D5 and D8 as a function of the JoFET dimensions is not fully understood. For these reasons, the $I_{\textrm{c}} R_{\textrm{N}}$ of such JoFET devices is not sufficient to characterize the quality of the proximity effect, especially because the switching current probability distribution is rather broad at low temperatures~\cite{haxell2022large}.

\begin{figure*}[h!]
  \includegraphics[]{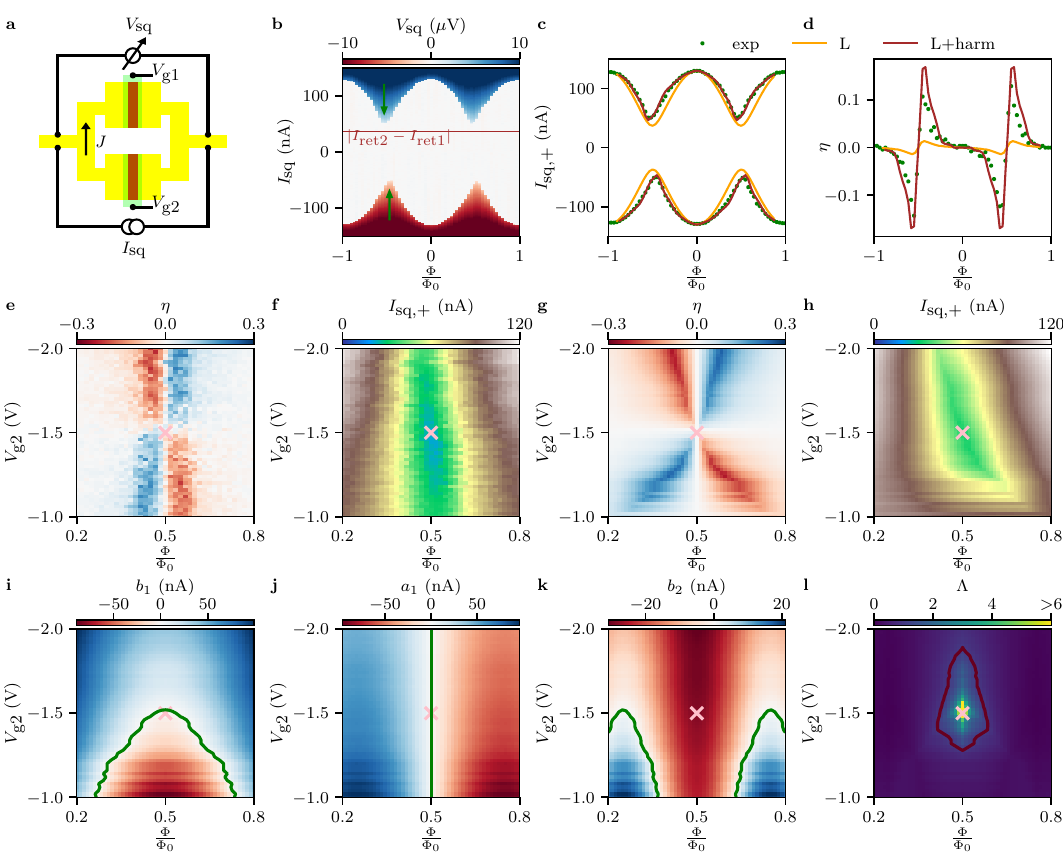} \caption{\textbf{Gate and flux tunable SDE employed as a generator of complex CPRs.} \small{ \textbf{a} Schematic of a typical SQUID. \textbf{b} $V_{\textrm{sq}}$ as a function of magnetic flux $\Phi$ and $I_{\textrm{sq}}$. $I_{\textrm{sq}}$ was swept from positive values to zero and, subsequently, from negative values to zero.  The SQUID lobes for positive and negative currents are asymmetric with respect to half flux and the critical current at $\frac{\Phi}{\Phi_0} = 0.5$ is larger than $\left| I_{\textrm{ret2}} - I_{\textrm{ret1}} \right|$, see the brown line. \textbf{c} Positive ($I_{\textrm{sq,+}}$) retrapping current extracted from \textbf{b}.  \textbf{d} Diode efficiency obtained from \textbf{b}.  $\eta = 0$ for integer and half-integer fluxes, whereas it reaches its maxima and minima around $\frac{\Phi}{\Phi_0} = 0.5$. The orange traces represent the expected outcome if the SQUID would be composed of standard tunnel junctions and it would have a total inductance of $L = \qty{110}{pH}$. The mere addition of $L$ is not enough to match the experimental data. The red traces represent the result of the numerical fit ($K_1=0.658,K_2=0.122,K_3=0.102$ and higher order terms are less than 10\%), imposing $L = \qty{110}{pH}$.  \textbf{e} [\textbf{f}] $\eta$ [$I_{\textrm{sq,+}}$]  as a function of $\Phi$ and $V_{\textrm{g2}}$ with $V_{\textrm{g1}} = \qty{-1.5}{V}$.  The behavior for the switching current is shown in the Extended Fig.~\ref{FigureSI_more_diode}. $\eta$ is always zero at $ \frac{\Phi}{\Phi_0} = 0.5$, independently on $V_{\textrm{g2}}$, and the polarity of the diode is inverted at the balanced point (pink cross). \textbf{g} [\textbf{h}] Theoretical calculation  of \textbf{e} [\textbf{f}], showing qualitatively similar behavior like the measurements.  \textbf{i-j} First harmonic contribution extracted from \textbf{g}. The green traces highlight the points where  their contribution vanishes. Just at the balanced point and at $\frac{\Phi}{\Phi_0} = 0.5$ both terms vanish.  \textbf{k} Second harmonic contribution; it never vanishes at $\frac{\Phi}{\Phi_0} = 0.5$. The cosinusoidal contribution is shown in Fig.~\ref{figure_SI_harmonics} along with higher order terms. \textbf{l} Ratio between second and first harmonic, $\Lambda = \frac{|b_2| + |a_2|}{|b_1| + |a_1|}$ as a function of $V_{\textrm{g2}}$ and $\Phi$. The red trace indicates the points where $K=1$. In \textbf{c}, \textbf{d} and \textbf{g-l}, $K_1=0.65,K_2=0.15,K_3=0.1$ and higher order terms are less than 10\% .}} 
\label{figure4}
\end{figure*}

\section*{Tunability of the induced superconducting gap}

The above reported results demonstrate that proximity induced superconductivity can be achieved in Ge without a direct contact to the superconductor, therefore avoiding metallization issues \cite{reeg2018metallization}. However, it is not clear to what extent the ${\textrm{Si}}_{0.3}{\textrm{Ge}}_{0.7}$ spacer of thickness D is influencing the value of the induced superconducting gap $\Delta^*$ and the subgap density of states. We expect that $\Delta^*$ depends on the coupling $t$ between the Al and the QW, see the sketch of Fig.~\ref{figure2}\textbf{a}. Since ${\textrm{Si}}_{0.3}{\textrm{Ge}}_{0.7}$ acts as a tunnel barrier, $t$ should be strongly dependent on D. In other words, if D is very thin, we expect $\Delta^*$ to be similar to the parent gap of Al $\Delta$; whereas if D is very thick the two layers will be very decoupled (small $t$) and $\Delta^*$ will be quenched.

In order to investigate the dependence of $\Delta^*$ on D, tunneling spectroscopy experiments were performed to estimate the local density of states of the hybridized Ge QW, see Figs.~\ref{figure2}\textbf{b} and \textbf{c} for the experimental layout.

The differential conductance $dI/dV$, plotted in logarithmic scale, as a function of source-drain bias $V$ and fingergates voltage $V_{\textrm{dg}}$ for D8 is shown in Fig.~\ref{figure2}\textbf{d}. $dI/dV$ is suppressed symmetrically around $V = 0$ independently on $V_{\textrm{dg}}$ (Fig.~\ref{figure2}\textbf{e}); this signals the presence of a superconducting gap in the hybridized Ge hole gas. Interestingly, a double peak structure is revealed. The first peak appears at $V \approx \qty{240}{\mu eV}$ and the second at  $V \approx \qty{80}{\mu eV}$. We interpret them as the parent gap of Al, $\Delta$, and the induced gap in the Ge hole gas $\Delta^{*}$. Their presence is more evident if $dI/dV$ is plotted in a linear scale, like in Fig.~\ref{figure2}\textbf{f} where the line-cuts have been shifted vertically for the sake of clarity. The sub-gap conductance is suppressed by one order of magnitude compared to the above-gap value. 

\

We now turn our attention to the $dI/dV$ of sample D5 \ (Fig.~\ref{figure2}\textbf{g}). The difference between Fig.~\ref{figure2}\textbf{g} and~\ref{figure2}\textbf{d} is striking. The region of suppressed conductance around $V=0$ is larger in Fig.~\ref{figure2}\textbf{g}, demonstrating that the induced gap is bigger. The line-cuts, see Fig.~\ref{figure2}\textbf{h}, showcase a difference between the normal-state conductance and the conductance at $V = 0$ of about two orders of magnitude, indicative of a hard gap \cite{krogstrup2015epitaxy}. Also for this device the double peak structure is observed (Fig.~\ref{figure2}\textbf{h} and Fig.~\ref{figure2}\textbf{i}). The parent gap appears at a similar value, namely $\Delta \approx \qty{230}{\mu eV}$, while $\Delta^{*} \approx \qty{150}{\mu eV}$.

\section*{Superconducting diode effect}

Having demonstrated a hard superconducting gap, we use the hybrid Al/Ge platform, to build a  SQUID device which acts as a gate/flux tunable superconducting diode \cite{ando2020observation} and as a generator of non-sinusoidal current-phase relations (CPRs). The superconducting diode effect (SDE) can appear in a simple SQUID either if its inductance $L$ is significant \cite{fulton1972quantum,paolucci2023gate} or if the CPRs of the single junctions have higher order contributions, arising from Andreev bound states in a semiconducting junction \cite{souto2022josephson,legg2023parity} or from junctions in the dirty limit~\cite{lambert1997boundary,galaktionov2000second}.  

Fig.~\ref{figure4}\textbf{a} shows the schematics of a SQUID, with the underlying 4-probe current-biased electrical circuit. $I_{\textrm{sq}}$ is the current passing through the SQUID, $J$ is the current circulating in the SQUID and $V_{\textrm{sq}}$ is the measured voltage drop across the device. In the following, we always sweep $I_{\textrm{sq}}$ from positive/negative values to $I_{\textrm{sq}}=0$, such that the retrapping current ($I_{\textrm{sq,+(-)}}$) is recorded for both branches. The use of the retrapping current avoids the challenges arising from the stochastic nature of the switching current ~\cite{haxell2022large}.
Top gate voltages $V_{\textrm{g1}}$ and $V_{\textrm{g2}}$ are used to tune the retrapping current $I_{\textrm{ret1}}$ of JJ1 and $I_{\textrm{ret2}}$ of JJ2. We first tune the device to be slightly unbalanced, i.e. $I_{\textrm{ret1}} = \qty{46.5}{nA} \neq I_{\textrm{ret2}} = \qty{83.5}{nA}$, see methods for understanding how $I_{\textrm{ret1}}$ and $I_{\textrm{ret2}}$ have been determined for the SQUID geometry. Fig.~\ref{figure4}\textbf{b} represents a SQUID measurement for such configuration, where $V_{\textrm{sq}}$ is recorded as a function of $I_{\textrm{sq}}$ and $\Phi$. $I_{\textrm{sq,+ (-)}}$ is periodically modulated by $\Phi$, as can be clearly seen in Fig.~\ref{figure4}\textbf{c}. However, two features are observed, which are not expected for a negligible inductance SQUID composed by tunnel junctions. First, the retrapping current at $\Phi =\frac{\Phi_0}{2}$ is expected to be $| I_{\textrm{ret1}} - I_{\textrm{ret2}} | = \qty{37}{nA} $ (see brown horizontal line in Fig.~\ref{figure4}\textbf{b}), instead the measured value is around $\qty{52}{nA}$.  Moreover, the SQUID pattern is not symmetric with respect to $\Phi =\frac{\Phi_0}{2}$, see green arrows in Fig.~\ref{figure4}\textbf{b}. This asymmetry gives rise to a finite SDE. The diode efficiency, defined as $\eta = \frac{I_{\textrm{sq,+}}-|I_{\textrm{sq,-}}|}{I_{\textrm{sq,+}}+|I_{\textrm{sq,-}}|}$, is shown in Fig.~\ref{figure4}\textbf{d}.
In particular, $\eta = 0$ at integer ($\Phi = n \Phi_0$) and it changes its sign around $\Phi = \frac{n}{2} \Phi_0$. The maximum value observed for this device is around $15 \%$.

In order to understand these results, we solve the static equation of the system:
\begin{equation}
\begin{split}
\frac{I_{\textrm{sq}}}{2} + J = I_{\textrm{JJ1}}\left( \varphi_1 \right), \\
\frac{I_{\textrm{sq}}}{2} - J = I_{\textrm{JJ2}}\left( \varphi_2 \right).
\end{split}
\label{eq:squid}
\end{equation}
$I_{\textrm{JJ1}}$ [$I_{\textrm{JJ2}}$] is the current flowing through JJ1 [JJ2] which depends on the phase difference across the junction $\varphi_1$ [$\varphi_2$]. The phase drops are related to the fluxoid quantization:

\begin{equation}
\varphi_2-\varphi_1 = 2 \pi \frac{\Phi}{\Phi_0} + 2 \pi \frac{L J}{\Phi_0}
\label{eq:phase}
\end{equation}
For a given $\Phi$, $I_{\textrm{sq,+}}$ [$I_{\textrm{sq,-}}$] is obtained by finding the maximum [minimum] $I_{\textrm{sq}}$ with  respect to $\varphi_1$. 

First, we attempt to understand our results assuming standard sinusoidal CPRs, i.e. $I_{\textrm{JJ1}}\left( \varphi_1 \right) = I_{\textrm{ret1}} \sin \left( \varphi_1 \right)$ and $I_{\textrm{JJ2}}\left( \varphi_1 \right) = I_{\textrm{ret2}} \sin \left( \varphi_2 \right)$, and adding the inductive contribution. $L$ is composed of two terms, a geometric one $L_{\textrm{geo}}$ and a kinetic one $L_{\textrm{kin}}$; we {extracted $L = \qty{110}{pH}$, see methods for details. 
The orange traces in Figs.~\ref{figure4}\textbf{c}-\textbf{d} represent the theoretical prediction. It is clear that the mere addition of a realistic $L$ does not capture the full picture, especially around $\frac{\Phi}{\Phi_0} = 0.5$, where $I_{\textrm{sq,+}}$ and $| I_{\textrm{sq,-}} |$ are greatly underestimated, see Fig.~\ref{FigureSI_L_harm_dependence}.

Therefore, it is necessary to consider higher order harmonics for explaining our results, namely we assume that our single junction CPRs are given by:

\begin{equation}
\begin{split}
I_{\textrm{JJ1}}\left( \varphi_1 \right) = \alpha_1 I_{\textrm{ret1}} \sum_n  {\left( -1 \right)}^{n+1} K_n \sin \left( n \varphi_1 \right)\\
I_{\textrm{JJ2}}\left( \varphi_2 \right) = \alpha_2 I_{\textrm{ret2}} \sum_n  {\left( -1 \right)}^{n+1} K_n \sin \left( n \varphi_2 \right)\\,
\label{eq:cpr}
\end{split}
\end{equation}
where $K_n$ is the relative contribution of the n-th harmonic and we assumed that the harmonics' contribution is the same for both junctions. $\alpha_1$ [$\alpha_2$] is a dimensionless parameter which is adjusted such that $\max{I_{\textrm{JJ1}}} = I_{\textrm{ret1}}$ [$\max{I_{\textrm{JJ2}}} = I_{\textrm{ret2}}$].

The red traces in Figs.~\ref{figure4}\textbf{c}-\textbf{d} are the outcome of a numerical fit using up to eight harmonic contributions. It is found that $K_{1}=0.66$,  $K_{2}=0.12$ and $K_{3}=0.10$, while higher order terms are smaller than $10 \%$, see Fig.~\ref{FigureSI_L_harm_dependence} to understand the effect of higher order terms.  We point out that also asymmetric cases would give qualitatively similar results (Fig.~\ref{figureSI_asymmetrysmall}).but the amount of free parameters of the fit would increase considerably. 

 We now turn our attention to the gate dependence of the SDE. Our measurements show that the SDE can be tuned by the gate voltages $V_{\textrm{g1}}$ and $V_{\textrm{g2}}$. In Fig.~\ref{figure4}\textbf{e}, we fix  $V_{\textrm{g1}} = \qty{-1.5}{V}$ and we study the behavior of $\eta$ while varying $\Phi$ and $V_{\textrm{g2}}$. When $ \left| V_{\textrm{g2}} \right| > \left| V_{\textrm{g1}} \right|$, $\eta > 0$ [$\eta < 0$] for $\frac{\Phi}{\Phi_0} > 0.5$ [$\frac{\Phi}{\Phi_0} < 0.5$], while the trend is opposite if $ \left| V_{\textrm{g2}} \right| < \left| V_{\textrm{g1}} \right|$ . In other words, we have an inversion of the diode polarity going from one regime to the other and most importantly, the SDE completely vanishes independently of $\Phi$ when the two junctions are fully balanced ($V_{\textrm{g1}} \approx V_{\textrm{g2}}$, i.e. $I_{\textrm{ret1}}=I_{\textrm{ret2}}=I_{\textrm{ret}}$).  
 
 $I_{\textrm{sq,+}}$, plotted in Fig.~\ref{figure4}\textbf{f}, does not vanish even at half flux quantum and for balanced junctions, see the pink cross in Fig.~\ref{figure4}\textbf{f}; we refer to this condition as the sweet spot. This is a crucial aspect, because at the sweet spot, the first harmonic of  $I_{\textrm{sq}}$ ($\propto \sin\left( \varphi \right)$) is completely suppressed but not the higher-order terms. We can understand this from considering Eqs.~\ref{eq:squid} at the sweet spot, 

\begin{equation}
\begin{split}
\frac{I_{\textrm{sq}}}{2} + J = \alpha K_1 I_{\textrm{ret}} \sin \left( \varphi_1 \right) -  \alpha K_2 I_{\textrm{ret}} \sin \left( 2 \varphi_1 \right) \\
\frac{I_{\textrm{sq}}}{2} - J =  \alpha K_1 I_{\textrm{ret}} \sin \left( \varphi_1 + \pi \right) - \alpha K_2 I_{\textrm{ret}} \sin \left( 2 \varphi_1 + 2 \pi \right) ,
\end{split}
\label{eq:squid_balanced}
\end{equation}
where for simplicity the inductance and higher order terms are neglected and $\alpha_1=\alpha_2=\alpha$. Therefore the CPR of the SQUID would be $I_{\textrm{sq}} \left( \varphi_1 \right) = - 2 \alpha  K_2 I_{\textrm{ret}} \sin \left( 2 \varphi_1 \right)$. This CPR corresponds to transport through the SQUID being governed by pairs of Cooper pairs, while the exchange of single pairs is quenched. This is the condition needed for creating a certain type protected qubit~\cite{smith2020superconducting}.

This behaviour can be further understood by solving Eqs.\ref{eq:squid}, \ref{eq:phase} and assuming to have JJs with higher order contributions. Fig.~\ref{figure4}\textbf{g} [\textbf{h}] represents the theoretical calculation of $\eta$ [$I_{\textrm{sq,+}}$] for a SQUID with the parameters extracted from the fit of Figs.~\ref{figure4}\textbf{c-d}. The simulation results agree well on a qualitative level with the measurements.
From the theoretical calculation it is possible to calculate the CPR of the SQUID and express it as a Fourier expansion:

\begin{equation}
\begin{split}
I_{\textrm{sq}} \left( \varphi \right) \approx b_1 \sin \left( \varphi \right) + b_2 \sin \left(2  \varphi \right) + \dots\\
+ a_1 \cos \left( \varphi \right) +  a_2 \cos \left( 2\varphi \right) + \dots
\end{split}
\label{eq:squidCPR}
\end{equation}
where $b_{\textrm{n}}$ and $a_{\textrm{n}}$ represents the n-th harmonic contribution and $\varphi$ is the phase drop across the SQUID. The values of the first harmonic terms $b_1$,  $a_1$ and second harmonic term $b_2$ obtained from numerical simulations are shown in  Figs.~\ref{figure4}\textbf{i-k}.

 Finally, we show the theoretical prediction of the ratio between the second and first harmonic, i.e.  $K = \frac{|b_2| + |a_2|}{|b_1| + |a_1|}$, see Fig.~\ref{figure4}\textbf{l}. The red trace depicts the points where the first and second harmonics equally contribute to the SQUID CPR, whereas the ratio diverges close to the sweet spot.

We note that, different CPRs of the single Josephson junctions would give slightly different outcomes. However, it would not change the main conclusion that $b_1$  and $a_1$ can be completely suppressed.  Moreover, the first harmonic contribution can be suppressed over a broad range of gate space, which also allows to tune the second harmonic contribution (Fig.~\ref{FigureSI_gatevsgate}). 

Finally, we note that the first harmonic can be quenched by just having a high inductance and the possibility of tuning the critical currents, see Fig.~\ref{fig:reply2}.

\begin{figure*}[h!]
  \includegraphics[]{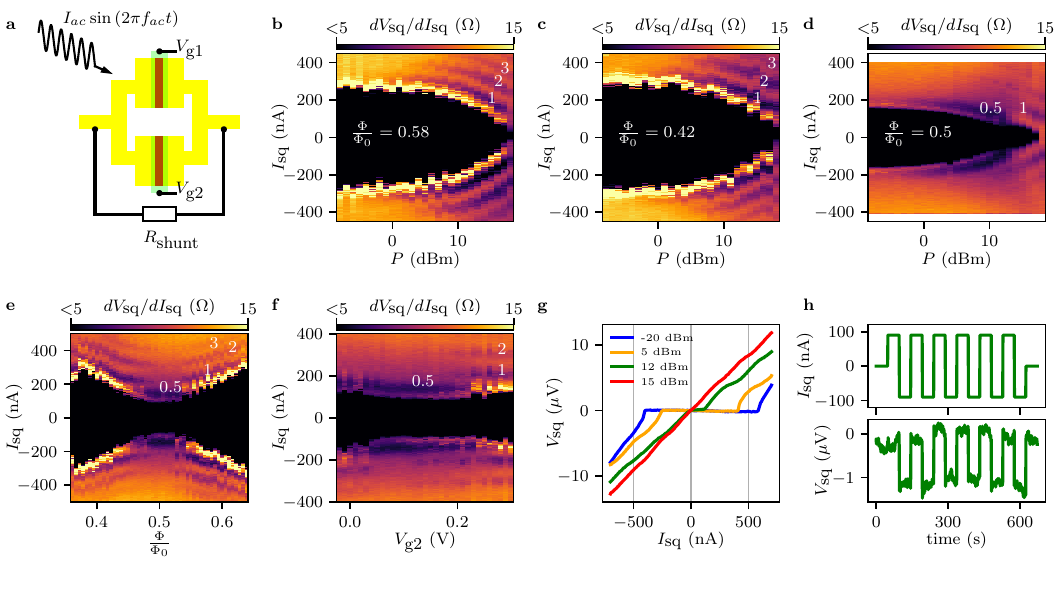}  \caption{\textbf{Half-integer Shapiro steps and ideal superconducting diode. } \small{ \textbf{a} Schematics of a typical SQUID used for Shapiro experiments with a microwave drive $I_{\textrm{ac}} \sin \left( 2 \pi f_{\textrm{ac}} t \right)$. \textbf{b} [\textbf{c}] Shapiro pattern for a $\qty{30}{nm}$ thick room temperature deposited Al sample  with $R_{\textrm{shunt}} = \qty{20}{\Omega}$ in the balanced regime ($I_{\textrm{ret1}} =I_{\textrm{ret2}} = \qty{500}{nA}$)   with $f_{\textrm{ac}} = \qty{500}{MHz}$ and at $\frac{\Phi}{\Phi_0} = 0.58$ [$\frac{\Phi}{\Phi_0} = 0.42$]. The differential resistance $dV_{\textrm{sq}}/dI_{\textrm{sq}}$ is plotted as a function of the RF power $P$ and $I_{\textrm{sq}}$. Dips in $dV_{\textrm{sq}}/dI_{\textrm{sq}}$ correspond to integer Shapiro steps. \textbf{d} same measurement as \textbf{b} and \textbf{c} but at $\frac{\Phi}{\Phi_0}=0.5$. Importantly at half flux quantum, the first half-integer steps appear for low $P$. \textbf{e} Shapiro map as a function of $I_{\textrm{sq}}$ and $\Phi$ in the balanced regime (like for the previous plots) for $P = \qty{9}{dBm}$. The half-integer steps appear only close to $\frac{\Phi}{\Phi_0}=0.5$. \textbf{f} Shapiro map as a function of $I_{\textrm{sq}}$ and $V_{\textrm{g2}}$ for $P = \qty{9}{dBm}$  and at $\frac{\Phi}{\Phi_0}=0.5$. The half-integer steps appear when the SQUID is close to the balanced condition, i.e. if $I_{\textrm{ret2}} \approx 500 \pm 40 \unit{nA}$. \textbf{g} and \textbf{h} show data from another $\qty{30}{nm}$ thick Al sample but at $f_{\textrm{ac}} = \qty{2}{GHz}$ and with  $R_{\textrm{shunt}} = \qty{50}{\Omega}$ for achieving a better visibility. \textbf{g} Current-voltage characteristic for $\Phi= 0.39 \Phi_0$  and for $f_{\textrm{ac}} = \qty{2} {GHz}$ with $I_{\textrm{ret1}} = \qty{670}{nA} $ and $I_{\textrm{ret2}} = \qty{450}{nA}$ for different power $P$.  At $P = \qty{12}{dBm}$, $I_{\textrm{sq,+}}$ is $\qty{120}{nA}$, whereas $I_{\textrm{sq,-}} \approx 0$, i.e. $\eta \approx 1$. \textbf{h} The non-volatility of SDE at $P = \qty{12}{dBm}$ is demonstrated by switching between the normal and superconducting behaviour alternating $I_{\textrm{sq}}$ from $\qty{90}{nA}$ to $\qty{-90}{nA}$ [upper panel]. In the lower panel, the measured voltage $V_{\textrm{sq}}$ is reported. A time dependent offset of $V_{\textrm{sq}}$, due to drift, was subtracted.}}
  \label{figure5}
\end{figure*}

\section*{Half-integer Shapiro steps and ideal SDE}

The good qualitative match between the experiment and the theoretical prediction of Fig.~\ref{figure4} makes us confident in our interpretation of the diode data, but the CPR of the SQUID was not directly probed. The AC Josephson effect would help to further elucidate the CPR periodicity. In fact, for a standard sinusoidal CPR under microwave irradiation, the current-voltage characteristics develop voltage steps when $V = s \frac{h f_{\textrm{ac}}}{2 e}$, the so-called Shapiro steps, where $s = 0,1,2,\dots$ and $f_{\textrm{ac}}$ is the external applied frequency. On the contrary, if the CPR becomes $\propto \sin \left(2  \varphi \right)$, signalling tunneling of pairs of Cooper pairs, steps at half-integer values also appear, i.e. $s = 0,0.5,1,\dots$ \cite{ueda2020evidence,zhang2022large,iorio2023half}. In our case, we expect the ratio between the second and first harmonic to be maximized when the SQUID is balanced and $\Phi \approx \frac{\Phi_0}{2}$, see Fig.~\ref{figure4}\textbf{l} as an example. Therefore, we would expect to observe half-integer Shapiro steps when approaching the sweet spot \cite{souto2022josephson}.

In order to avoid flux generated by inductive effects which might lead to similar results \cite{vanneste1988shapiro}, we present results of a 30 nm-thick aluminum SQUID, which has a much smaller inductance ($L < \qty{15}{pH}$). Furthermore, a shunt resistor $R_{\textrm{shunt}}$ of 10-50 $\unit{\Omega}$, see Fig.~\ref{figure5}\textbf{a}, was added in order to create overdamped junctions, allowing therefore to measure Shapiro steps at small external frequencies, avoiding issues related to Landau-Zener transitions \cite{dartiailh2021missing}.

In the following, we study a SQUID in a balanced configuration ($I_{\textrm{ret1}} = I_{\textrm{ret2}}$) subjected to an external drive at $f_{\textrm{ac}} = \qty{500}{MHz}$. Fig.~\ref{figure5}\textbf{b} shows the differential resistance of the SQUID $dV_{\textrm{sq}}/dI_{\textrm{sq}}$ as a function of the microwave drive power $P$ and $I_{\textrm{sq}}$ at $\frac{\Phi}{\Phi_0}=0.58$. If $P$ is high enough, dips corresponding to the integer Shapiro steps $s = 1,2,3$ appear. Similar results are obtained at $\frac{\Phi}{\Phi_0}=0.42$, see Fig.~\ref{figure5}\textbf{c}. However, the situation is different if $\frac{\Phi}{\Phi_0}=0.5$ (Fig.~\ref{figure5}\textbf{d}) a condition for which the first harmonic term should vanish. For this situation, the first half-integer steps appears as theoretically expected (see Fig.~\ref{figureSI_shapiro_10Ohm} for the identification of the Shapiro steps).  

This behavior is summarized in Fig.~\ref{figure5}\textbf{e} where we fix $P = \qty{9}{dBm}$ and we display $dV_{\textrm{sq}}/dI_{\textrm{sq}}$ as a function of $I_{\textrm{sq}}$ and $\Phi$. Far from half flux quantum, dips corresponding to integer Shapiro steps are observed; while close to $\frac{\Phi}{\Phi_0} =0.5$ the integer steps fade and the $s=0.5$ step becomes pronounced, see white numbers. In order to further investigate the range over which the half-integer Shapiro step is visible, we fix $\frac{\Phi}{\Phi_0} =0.5$ and we vary $I_{\textrm{ret2}}$ with the gate voltage $V_{\textrm{g2}}$. When the SQUID is close to the balanced regime ($V_{\textrm{g2}} \approx \qty{0.1}{V}$) the first half-integer step is evident (white numbers); however it fades away if the SQUID becomes unbalanced, i.e. $V_{\textrm{g2}} > \qty{0.2}{V}$. As expected from the previous analysis, the half-integer step appears only if the device is close to the balanced position and close to half flux quantum when $I_{\textrm{sq}} \left( \varphi_1 \right) =b_2 \sin \left( 2 \varphi_1 \right) + b_4 \sin \left( 4 \varphi_1 \right) + \dots$. Importantly, also a  second device investigated under microwave irradiation showed the same behavior, see Fig.~\ref{figureSI_shapiro_10Ohm}.  

The SDE indicates that the symmetry $I_{\textrm{sq,+}} = -I_{\textrm{sq,-}}$  is broken in our system, which also implies that the widths $\Delta I_{\pm 1}$ of the two first Shapiro steps, which eventually define $I_{\textrm{sq,+}}$ and $I_{\textrm{sq,-}}$, can be different ~\cite{souto2022josephson}. As the position and the width of the plateaus depend on the microwave drive, one can envision tuning to a situation in which the first negative plateau would start at zero current ($I_{\textrm{sq,-}}=0$) while the first positive one at a finite current ($I_{\textrm{sq,+}}\neq0$). At this particular strength of the ac driving, the SQUID is  expected to become an ideal SDE, i.e. $\eta \approx 1$. 


In order to investigate this possibility a similar SQUID, with $R_{\textrm{shunt}} = \qty{50}{\Omega}$,  at $\Phi = 0.39 \Phi_0$ for different drive powers $P$ was investigated (Fig.~\ref{figure5}\textbf{g}). 
 For small $P$, $I_{\textrm{sq,+}} > |I_{\textrm{sq,-}}|$ and $\eta \approx 0.18$, see blue trace. When $P$ increases both $I_{\textrm{sq,+}}$ and $|I_{\textrm{sq,-}}|$ decrease, see orange trace. Eventually when  $P$ is high enough (green trace), $I_{\textrm{sq,-}}$ drops to zero, whereas $I_{\textrm{sq,+}} \neq 0$, yielding a diode efficiency equal to 1.  Moreover, we show that our device is non-volatile, namely we can switch several times from the normal-state to the superconducting branch by changing the current direction, see Fig.~\ref{figure5}\textbf{h}.

\begin{figure*}[b]
  \includegraphics[]{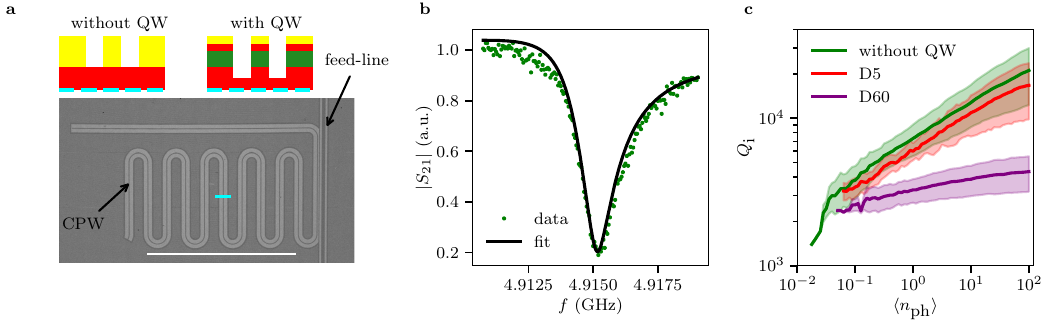}  \caption{\textbf{Hybrid coplanar waveguide resonators}. \small{ \textbf{a} Scanning electron microscopy image of a typical notch-type CPW resonator. The top insets show sketches of the CPW cross-section, at the position of the cyan trace, when the QW has been etched away (left) and when the resonator is fabricated on top of the QW (right). In the latter case, the feedline, the resonator and the ground plane are formed on top of the Ge QW. The scale bar corresponds to $500 \ \si{\mu m}$.\ \textbf{b} $\left| S_{21} \right|$ signal as a function of the frequency $f$ around the resonance frequency of a device without QW. The black line is an algebraic fit~\cite{probst2015efficient}. \textbf{c} Internal quality factor $Q_{\textrm{i}}$, averaged for multiple  $\qty{4}{GHz} < f_{\textrm{r}} < \qty{5.5}{GHz}$ resonators (Table~\ref{tab:resonator}), as a function of the average photon number $\langle n_{\textrm{ph}} \rangle $ for  samples without QW, D5 and D60.  The shaded areas correspond to the standard deviation of the data. We notice that the proximitized Ge QW in the samples D5 leads to small additional losses, compared to the without QW resonator samples. However, if the Ge QW gets decoupled from the Al (D60), we observe a lower $Q_{\textrm{i}}$, suggesting that $Q_{\textrm{i}}$ is limited by losses in Ge.}}
  \label{figure3}
\end{figure*}

\section*{Discussion/Outlook}
In the past few years, planar germanium has established itself as a promising platform for spin-qubit arrays~\cite{scappucci2021}. Here, we demonstrate its potential also for hybrid semiconductor-superconductor quantum devices. Inspired by more mature technologies \cite{Mayer2019}, we introduced a reliable way to induce superconductivity  by using shallow QWs and, to the best of our knowledge, we have realized the largest hard gap in Ge. Our method does not rely on the precise etching of the QW and/or surface treatments \cite{aggarwal2021enhancement}  and does not require in-situ deposition of the superconductor. Furthermore, it minimizes the Fermi velocity mismatch due to the direct contact between Ge and proximitized Ge, enhancing Andreev reflection over normal reflection. 

While the shallow QWs reported in this work are of limited mobility and have a larger charge noise, which can be a challenge for the realization of scalable spin qubits, possible mitigation strategies of this problem could include a careful engineering of the semiconductor/dielectric interface \cite{DegliEsposti2022}, including the use of Ge caps~\cite{Su}, or growing the QWs on Ge instead of Si wafers~\cite{stehouwer2023germanium}. A further solution could be to have a thin spacer in the areas where superconductivity should be induced and a thicker one in the areas where the spin qubits will be formed. 

 The reported large superconducting hard gap on a group IV material will enable spin qubit coupling via coherent tunneling and cotunneling processes that involve (crossed) Andreev reflection~\cite{Leijnse2013,spethmann2023}. In addition, the realized gate and flux-tunable superconducting diode can  suppress the first harmonic term, making it therefore an interesting building block for creating protected superconducting qubits with semiconductor materials \cite{brooks2013protected,smith2020superconducting,schrade2022protected,larsen2020parity,gyenis2021experimental}. In order to realize such qubits, superconducting resonators are key elements. A $\lambda / 4$ notch-type resonator is shown in (Fig.~\ref{figure3}\textbf{a}) (see methods for details). The upper-left inset of Fig.~\ref{figure3}\textbf{a} depicts the cross-section of the  resonator, pointing out that the Ge QW has been completely etched away prior to the resonator fabrication. Fig.~\ref{figure3}\textbf{b} shows the transmission amplitude $\left| S_{21} \right|$ as a function of the probe frequency $f$. The internal quality factors $Q_{\textrm{i}}$ were extracted  \cite{probst2015efficient} and found to be around 7000 [20000] for $ \langle n_{\textrm{ph}} \rangle \approx 1$ [100]; demonstrating the microwave compatibility of the used Ge/SiGe heterostructures. Interestingly, just slightly smaller $Q_{\textrm{i}}$ values were extracted also for superconducting resonators fabricated on Ge/SiGe heterostructures where the Ge QW has been removed just in the gap between the central conductor and the ground plane, showing that the proximitized Ge does not lead to significant losses. 
The above demonstrated microwave compatibility of the used Ge/SiGe heterostructures opens a path towards spin-photon experiments \cite{Burkard2020}, gate tunable transmon qubits~\cite{larsen2015semiconductor,de2015realization,casparis2018superconducting} and superconducting spin qubits in group IV materials~\cite{pita2022direct} and allow us to envision the transfer of quantum information between different types of qubits, all realizable on planar Ge.

After submission of our manuscript we became aware of similar works dealing with the superconducting diode effect in interferometer devices~\cite{ciaccia2023charge,li2023interfering,matsuo2023josephson}.

\clearpage

\section*{Methods}
\subsubsection*{Growth and Al deposition}
Strained Ge QW structures were grown by low-energy plasma-enhanced chemical vapor deposition on forward-graded buffers~\cite{jirovec2021singlet} with ${\textrm{Si}}_{0.3}{\textrm{Ge}}_{0.7}$ caps of 5 and 8 $\si{nm}$ above the 18 $\si{nm}$ Ge QW.  These nominal QW and cap thicknesses vary across the wafer due to the intensity profile of the focused plasma. Thicknesses were verified by comparing high-resolution x-ray diffraction $\omega$--2$\theta$ scans with dynamical simulations based on a smoothed QW profile~\cite{jirovec2021singlet}. This same composition profile was constructed within the \texttt{NextNano} 1-d Poisson-Schr\"odinger solver, along with a dieletric layer and top Schottky contact, in order to generate the band profile and wavefunction density shown in Fig.~\ref{figure1}\textbf{a}.
The Ge/SiGe heterostructures are cut in pieces of 6x6 $\unit{mm^2}$. A $\qty{3}{min}$ buffered oxide etch removes the native surface oxide of the diced samples, after which they are transferred to a molecular beam epitaxy chamber. The samples are cooled down to $\qty{110}{K}$ by active liquid-nitrogen cooling. Subsequently, Al is deposited at a growth rate of $\qty{5.5}{\AA \per min}$. Immediately after growth, samples are transferred in-situ to a chamber equipped with an ultrahigh-purity $\textrm{O}_2$ source where they are exposed to $10^{-4} \unit{mbar}$ of $\textrm{O}_2$ for $\qty{15}{min}$. The formed oxide layer prevents subsequent retraction of the metal film as the sample warms up to room temperature under ultrahigh vacuum conditions.

\subsubsection*{Sample fabrication}
\noindent \textbf{10 nm-thick, cold deposited aluminum samples:}\\
A mesa of around $\qty{60}{nm}$ depth is obtained by first removing Al with Transene D and then by etching the heterostructure with a ${\textrm{SF}}_{6}$-${\textrm{O}}_{2}$-${\textrm{CHF}}_{3}$ reactive ion etching process. In a second step, Al is selectively etched away using Transene D in order to create the Josephson junction or tunneling spectroscopy devices. Then, for tunneling spectroscopy devices, normal metal ohmic contacts are created by argon milling the SiGe spacer followed by a deposition of $\qty{60}{nm}$ platinum at an angle of 5\textdegree.  Finally, 9-18 nm plasma assisted aluminum oxide is deposited on top of all the sample at 150$^{\circ}$C and then Ti/Pd gates are evaporated. For some devices, two layers of top-gates were needed.

\noindent \textbf{30 nm-thick, room-temperature deposited aluminum samples:}\\
A mesa of around $\qty{60}{nm}$ depth is obtained by etching the heterostructure with a ${\textrm{SF}}_{6}$-${\textrm{O}}_{2}$-${\textrm{CHF}}_{3}$ reactive ion etching process. The sample is then submerged for $\qty{15}{s}$ in buffered HF and, subsequently, the $\qty{30}{nm}$ Al film is deposited. Gates are patterned like for the $\qty{10}{nm}$-thick sample. Importantly, this technique allows to fabricate devices without the need of wet etching for removing the superconductor.

\noindent \textbf{CPW resonator, without QW:}\\
The QW is removed by a ${\textrm{SF}}_{6}$-${\textrm{O}}_{2}$-${\textrm{CHF}}_{3}$ reactive ion etching process. Subsequently, the CPW resonator, the feed-line and the ground plane are written by electron beam lithography followed by a 25 nm-thick Al deposition at room temperature.

\noindent \textbf{CPW resonator, with QW:}\\
Electron beam lithography is performed on a sample with low temperature deposited Aluminum. The area between the ground plane and the signal line is exposed and after development the Al is removed by transene D etching. Finally, and before removing the resist the Ge QW is etched away by a ${\textrm{SF}}_{6}$-${\textrm{O}}_{2}$-${\textrm{CHF}}_{3}$ reactive ion etching process.

  \begin{table*}[htbp]
\centering
\begin{tabular}{|c|c|c|}
\hline
 & $\qty{10}{nm}$ Al & $\qty{30}{nm}$ Al \\
\hline
$R_{\square}$ & $ \qty{12.2}{\Omega}$ & $  \qty{0.75}{\Omega}$ \\
\hline
$T_{\textrm{c}}$ & $ \qty{1.9}{K}$ & $ \qty{1.4}{K}$ \\
\hline
$L_{\textrm{kin},\square}$ & $ \qty{8.9}{pH}$ & $ \qty{0.75}{pH}$ \\
\hline
$L_{\textrm{kin}}+L_{\textrm{geo}}$ & $ \qty{110}{pH}$ & $  \qty{13}{pH}$ \\
\hline
\end{tabular}
\caption{\textbf{Estimation of SQUID inductance}. \small{ $R_{\square}$ and $T_{\textrm{c}}$ were estimated from 4-probe current biased measurements. The SQUIDs are consisting of approximately 12 squares.}}
\label{tab:L}
\end{table*}

 \subsubsection*{Inductance estimation}
 The total inductance $L$ of the measured SQUIDs has two contributions: the geometric one ($L_{\textrm{geo}}$) and the kinetic one ($L_{\textrm{kin}}$). For the $L_{\textrm{geo}}$ we approximated the device to a loop with a radius $R$ and with the wire diameter $d$ 
 \begin{equation}
     L_{\textrm{geo}} = \mu_0  R \left[ \ln \left(\frac{16 R}{d} \right) -2
 \right]
 \end{equation}
 where $\mu_0$ is the magnetic vacuum permeability and we assumed the relative magnetic permeability $\mu_{\textrm{r}} =1$. Typical values of our SQUID geometry are $R \approx \qty{1.25}{\mu m}$ and $d \approx \qty{0.7}{\mu m}$, which gives $L_{\textrm{geo}}  \approx \qty{2}{pH}$. However, in order not to underestimate this contribution, $L_{\textrm{geo}}$ is assumed to be as large as $\approx \qty{5}{pH}$.
As regards the kinetic inductance per square $L_{\textrm{kin},\square}$, it was estimated from the values of the superconducting gap and the square normal-state resistance $R_{\square}$ \cite{annunziata2010tunable}:
 \begin{equation}
L_{\textrm{kin},\square} = \frac{h}{2 \pi^2}\frac{R_{\square}}{\Delta}
 \end{equation}
 where $\Delta$ is estimated from the critical temperature $T_{\textrm{c}}$, i.e.  $\Delta = 1.76 k_{\textrm{B}} T_{\textrm{c}}$ with $k_{\textrm{B}}$ being the Boltzmann constant. The results are summarized in Table~\ref{tab:L}. 

 \subsubsection*{Estimation of the retrapping currents in the SQUID geometry}
We note that the sum of the retrapping currents measured in isolation ($I_{\textrm{ret1,iso}}$ and $I_{\textrm{ret2,iso}}$), i.e. with the other junction pinched off, in the absence of a shunt resistor is always smaller than the retrapping current of the squid at $\Phi=0$. This difference is attributed to the fact that the SQUID has a smaller resistance, which leads to lower dissipation \cite{courtois2008origin} and, as a result, a higher retrapping current. Therefore, we assume an even redistribution of retrapping currents such that $\frac{I_{\textrm{ret1,iso}}}{I_{\textrm{ret2,iso}}} = \frac{I_{\textrm{ret1}}}{I_{\textrm{ret2}}}$ and $I_{\textrm{ret1}}+I_{\textrm{ret2}} = I_{\textrm{sq,+}} \left(\Phi=0\right)$. This approach was used to estimate the retrapping currents for Fig.~\ref{figure4}.}

 \subsection*{Acknowledgment} 
The authors acknowledge Alexander Brinkmann, Alessandro Crippa, Francesco Giazotto, Andrew Higginbotham, Andrea Iorio, Giordano Scappucci, Christian Schonenberger and Lukas Splitthoff for helpful discussions. We thank Marcel Verheijen for the support in the TEM analysis. This research and related results were made possible with the support of the NOMIS Foundation. It was supported by the Scientific Service Units of ISTA through resources provided by the MIBA Machine Shop and the nanofabrication facility, the European Union's Horizon 2020 research and innovation programme under Grant Agreement No 862046, the HORIZON-RIA 101069515 project, the HORIZON-EIC 101115315 project and the FWF Projects \#P-32235, \#P-36507 and \#F-8606. For the purpose of open access, the authors have applied a CC BY public copyright licence to any Author Accepted Manuscript version arising from this submission. R.S.S. acknowledges Spanish CM “Talento Program” Project No. 2022-T1/IND-24070. J.J. acknowledges European Research Council TOCINA 834290.


{\textbf{Author contribution}}. M.V. and T.G. fabricated the transport devices and performed the measurements under the supervision of G.K.. M.V. did the data analysis under the supervision of G.K.. L.B. and O.S. fabricated and measured the CPW. M.J. and O.S. developed the microwave technology for the Ge/SiGe heterostructures. K.A. contributed to the transport measurements and to the device fabrication. J.A.S. fabricated the Hall bars. T. A. made the Shapiro measurements possible. M.V. performed the simulations with input from C.S. and R.S.S.. S.C., A.B., D.C. and G.I. were responsible for the Ge QW growth, Hall bar measurements and NextNano simulations. J.J. and E.B. were responsible for the growth of low temperature Al and TEM data. M. L. and J. D. contributed to the interpretation of the experimental results. M.V. and G.K. wrote the manuscript with input from all the coauthors.

{\textbf{Competing interests}}: The authors declare no competing interests. 

{\textbf{Data and Materials Availability}}: All experimental data included in this work will be available at . 

\newpage

\bibliography{biblio.bib}

\begin{thebibliography}{10}
\expandafter\ifx\csname url\endcsname\relax
  \def\url#1{\texttt{#1}}\fi
\expandafter\ifx\csname urlprefix\endcsname\relax\def\urlprefix{URL }\fi
\providecommand{\bibinfo}[2]{#2}
\providecommand{\eprint}[2][]{\url{#2}}

\bibitem{krogstrup2015epitaxy}
\bibinfo{author}{Krogstrup, P.} \emph{et~al.}
\newblock \bibinfo{title}{Epitaxy of semiconductor--superconductor nanowires}.
\newblock \emph{\bibinfo{journal}{Nature materials}} \textbf{\bibinfo{volume}{14}}, \bibinfo{pages}{400--406} (\bibinfo{year}{2015}).

\bibitem{larsen2015semiconductor}
\bibinfo{author}{Larsen, T.~W.} \emph{et~al.}
\newblock \bibinfo{title}{Semiconductor-nanowire-based superconducting qubit}.
\newblock \emph{\bibinfo{journal}{Physical review letters}} \textbf{\bibinfo{volume}{115}}, \bibinfo{pages}{127001} (\bibinfo{year}{2015}).

\bibitem{de2015realization}
\bibinfo{author}{De~Lange, G.} \emph{et~al.}
\newblock \bibinfo{title}{Realization of microwave quantum circuits using hybrid superconducting-semiconducting nanowire josephson elements}.
\newblock \emph{\bibinfo{journal}{Physical review letters}} \textbf{\bibinfo{volume}{115}}, \bibinfo{pages}{127002} (\bibinfo{year}{2015}).

\bibitem{casparis2018superconducting}
\bibinfo{author}{Casparis, L.} \emph{et~al.}
\newblock \bibinfo{title}{Superconducting gatemon qubit based on a proximitized two-dimensional electron gas}.
\newblock \emph{\bibinfo{journal}{Nature nanotechnology}} \textbf{\bibinfo{volume}{13}}, \bibinfo{pages}{915--919} (\bibinfo{year}{2018}).

\bibitem{hays2021asq}
\bibinfo{author}{Hays, M.} \emph{et~al.}
\newblock \bibinfo{title}{Coherent manipulation of an andreev spin qubit}.
\newblock \emph{\bibinfo{journal}{Science}} \textbf{\bibinfo{volume}{373}}, \bibinfo{pages}{430--433} (\bibinfo{year}{2021}).

\bibitem{pita2022direct}
\bibinfo{author}{Pita-Vidal, M.} \emph{et~al.}
\newblock \bibinfo{title}{Direct manipulation of a superconducting spin qubit strongly coupled to a transmon qubit}.
\newblock \emph{\bibinfo{journal}{Nature Physics}} \bibinfo{pages}{1--6} (\bibinfo{year}{2023}).

\bibitem{phan2022semiconductor}
\bibinfo{author}{Phan, D.} \emph{et~al.}
\newblock \bibinfo{title}{Gate-tunable superconductor-semiconductor parametric amplifier}.
\newblock \emph{\bibinfo{journal}{Physical Review Applied}} \textbf{\bibinfo{volume}{19}}, \bibinfo{pages}{064032} (\bibinfo{year}{2023}).

\bibitem{wang2022singlet}
\bibinfo{author}{Wang, G.} \emph{et~al.}
\newblock \bibinfo{title}{Singlet and triplet cooper pair splitting in hybrid superconducting nanowires}.
\newblock \emph{\bibinfo{journal}{Nature}} \bibinfo{pages}{1--6} (\bibinfo{year}{2022}).

\bibitem{bordoloi2022spin}
\bibinfo{author}{Bordoloi, A.}, \bibinfo{author}{Zannier, V.}, \bibinfo{author}{Sorba, L.}, \bibinfo{author}{Sch{\"o}nenberger, C.} \& \bibinfo{author}{Baumgartner, A.}
\newblock \bibinfo{title}{Spin cross-correlation experiments in an electron entangler}.
\newblock \emph{\bibinfo{journal}{Nature}} \bibinfo{pages}{1--5} (\bibinfo{year}{2022}).

\bibitem{wang2022triplet}
\bibinfo{author}{Wang, Q.} \emph{et~al.}
\newblock \bibinfo{title}{Triplet cooper pair splitting in a two-dimensional electron gas}.
\newblock \emph{\bibinfo{journal}{arXiv preprint arXiv:2211.05763}}  (\bibinfo{year}{2022}).

\bibitem{dvir2022mini}
\bibinfo{author}{Dvir, T.} \emph{et~al.}
\newblock \bibinfo{title}{Realization of a minimal kitaev chain in coupled quantum dots}.
\newblock \emph{\bibinfo{journal}{Nature}} \textbf{\bibinfo{volume}{614}}, \bibinfo{pages}{445--450} (\bibinfo{year}{2023}).

\bibitem{ando2020observation}
\bibinfo{author}{Ando, F.} \emph{et~al.}
\newblock \bibinfo{title}{Observation of superconducting diode effect}.
\newblock \emph{\bibinfo{journal}{Nature}} \textbf{\bibinfo{volume}{584}}, \bibinfo{pages}{373--376} (\bibinfo{year}{2020}).

\bibitem{baumgartner2022supercurrent}
\bibinfo{author}{Baumgartner, C.} \emph{et~al.}
\newblock \bibinfo{title}{Supercurrent rectification and magnetochiral effects in symmetric josephson junctions}.
\newblock \emph{\bibinfo{journal}{Nature nanotechnology}} \textbf{\bibinfo{volume}{17}}, \bibinfo{pages}{39--44} (\bibinfo{year}{2022}).

\bibitem{pal2022josephson}
\bibinfo{author}{Pal, B.} \emph{et~al.}
\newblock \bibinfo{title}{Josephson diode effect from cooper pair momentum in a topological semimetal}.
\newblock \emph{\bibinfo{journal}{Nature physics}} \textbf{\bibinfo{volume}{18}}, \bibinfo{pages}{1228--1233} (\bibinfo{year}{2022}).

\bibitem{mazur2022gate}
\bibinfo{author}{Mazur, G.} \emph{et~al.}
\newblock \bibinfo{title}{The gate-tunable josephson diode}.
\newblock \emph{\bibinfo{journal}{arXiv preprint arXiv:2211.14283}}  (\bibinfo{year}{2022}).

\bibitem{steiner2023diode}
\bibinfo{author}{Steiner, J.~F.}, \bibinfo{author}{Melischek, L.}, \bibinfo{author}{Trahms, M.}, \bibinfo{author}{Franke, K.~J.} \& \bibinfo{author}{von Oppen, F.}
\newblock \bibinfo{title}{Diode effects in current-biased josephson junctions}.
\newblock \emph{\bibinfo{journal}{Physical Review Letters}} \textbf{\bibinfo{volume}{130}}, \bibinfo{pages}{177002} (\bibinfo{year}{2023}).

\bibitem{wu2022field}
\bibinfo{author}{Wu, H.} \emph{et~al.}
\newblock \bibinfo{title}{The field-free josephson diode in a van der waals heterostructure}.
\newblock \emph{\bibinfo{journal}{Nature}} \textbf{\bibinfo{volume}{604}}, \bibinfo{pages}{653--656} (\bibinfo{year}{2022}).

\bibitem{trahms2023diode}
\bibinfo{author}{Trahms, M.} \emph{et~al.}
\newblock \bibinfo{title}{Diode effect in josephson junctions with a single magnetic atom}.
\newblock \emph{\bibinfo{journal}{Nature}} \textbf{\bibinfo{volume}{615}}, \bibinfo{pages}{628--633} (\bibinfo{year}{2023}).

\bibitem{gupta2022superconducting}
\bibinfo{author}{Gupta, M.} \emph{et~al.}
\newblock \bibinfo{title}{Gate-tunable superconducting diode effect in a three-terminal josephson device}.
\newblock \emph{\bibinfo{journal}{Nature communications}} \textbf{\bibinfo{volume}{14}}, \bibinfo{pages}{3078} (\bibinfo{year}{2023}).

\bibitem{chilestriode}
\bibinfo{author}{Chiles, J.} \emph{et~al.}
\newblock \bibinfo{title}{Nonreciprocal supercurrents in a field-free graphene josephson triode}.
\newblock \emph{\bibinfo{journal}{Nano Letters}}  (\bibinfo{year}{2023}).

\bibitem{souto2022josephson}
\bibinfo{author}{Souto, R.~S.}, \bibinfo{author}{Leijnse, M.} \& \bibinfo{author}{Schrade, C.}
\newblock \bibinfo{title}{Josephson diode effect in supercurrent interferometers}.
\newblock \emph{\bibinfo{journal}{Physical Review Letters}} \textbf{\bibinfo{volume}{129}}, \bibinfo{pages}{267702} (\bibinfo{year}{2022}).

\bibitem{fominov2022asymmetric}
\bibinfo{author}{Fominov, Y.~V.} \& \bibinfo{author}{Mikhailov, D.}
\newblock \bibinfo{title}{Asymmetric higher-harmonic squid as a josephson diode}.
\newblock \emph{\bibinfo{journal}{Physical Review B}} \textbf{\bibinfo{volume}{106}}, \bibinfo{pages}{134514} (\bibinfo{year}{2022}).

\bibitem{ciaccia2023gate}
\bibinfo{author}{Ciaccia, C.} \emph{et~al.}
\newblock \bibinfo{title}{Gate-tunable josephson diode in proximitized {InAs} supercurrent interferometers}.
\newblock \emph{\bibinfo{journal}{Physical Review Research}} \textbf{\bibinfo{volume}{5}}, \bibinfo{pages}{033131} (\bibinfo{year}{2023}).

\bibitem{smith2020superconducting}
\bibinfo{author}{Smith, W.}, \bibinfo{author}{Kou, A.}, \bibinfo{author}{Xiao, X.}, \bibinfo{author}{Vool, U.} \& \bibinfo{author}{Devoret, M.}
\newblock \bibinfo{title}{Superconducting circuit protected by two-cooper-pair tunneling}.
\newblock \emph{\bibinfo{journal}{npj Quantum Information}} \textbf{\bibinfo{volume}{6}}, \bibinfo{pages}{8} (\bibinfo{year}{2020}).

\bibitem{brooks2013protected}
\bibinfo{author}{Brooks, P.}, \bibinfo{author}{Kitaev, A.} \& \bibinfo{author}{Preskill, J.}
\newblock \bibinfo{title}{Protected gates for superconducting qubits}.
\newblock \emph{\bibinfo{journal}{Physical Review A}} \textbf{\bibinfo{volume}{87}}, \bibinfo{pages}{052306} (\bibinfo{year}{2013}).

\bibitem{gyenis2021experimental}
\bibinfo{author}{Gyenis, A.} \emph{et~al.}
\newblock \bibinfo{title}{Experimental realization of a protected superconducting circuit derived from the 0--$\pi$ qubit}.
\newblock \emph{\bibinfo{journal}{PRX Quantum}} \textbf{\bibinfo{volume}{2}}, \bibinfo{pages}{010339} (\bibinfo{year}{2021}).

\bibitem{larsen2020parity}
\bibinfo{author}{Larsen, T.~W.} \emph{et~al.}
\newblock \bibinfo{title}{Parity-protected superconductor-semiconductor qubit}.
\newblock \emph{\bibinfo{journal}{Physical review letters}} \textbf{\bibinfo{volume}{125}}, \bibinfo{pages}{056801} (\bibinfo{year}{2020}).

\bibitem{schrade2022protected}
\bibinfo{author}{Schrade, C.}, \bibinfo{author}{Marcus, C.~M.} \& \bibinfo{author}{Gyenis, A.}
\newblock \bibinfo{title}{Protected hybrid superconducting qubit in an array of gate-tunable josephson interferometers}.
\newblock \emph{\bibinfo{journal}{PRX Quantum}} \textbf{\bibinfo{volume}{3}}, \bibinfo{pages}{030303} (\bibinfo{year}{2022}).

\bibitem{Maiani}
\bibinfo{author}{Maiani, A.}, \bibinfo{author}{Kjaergaard, M.} \& \bibinfo{author}{Schrade, C.}
\newblock \bibinfo{title}{Entangling transmons with low-frequency protected superconducting qubits}.
\newblock \emph{\bibinfo{journal}{PRX Quantum}} \textbf{\bibinfo{volume}{3}}, \bibinfo{pages}{030329} (\bibinfo{year}{2022}).

\bibitem{scappucci2021}
\bibinfo{author}{Scappucci, G.} \emph{et~al.}
\newblock \bibinfo{title}{{The germanium quantum information route}}.
\newblock \emph{\bibinfo{journal}{Nature Reviews Materials}} \textbf{\bibinfo{volume}{6}}, \bibinfo{pages}{926--943} (\bibinfo{year}{2021}).

\bibitem{xiang2006ge}
\bibinfo{author}{Xiang, J.}, \bibinfo{author}{Vidan, A.}, \bibinfo{author}{Tinkham, M.}, \bibinfo{author}{Westervelt, R.~M.} \& \bibinfo{author}{Lieber, C.~M.}
\newblock \bibinfo{title}{Ge/si nanowire mesoscopic josephson junctions}.
\newblock \emph{\bibinfo{journal}{Nature nanotechnology}} \textbf{\bibinfo{volume}{1}}, \bibinfo{pages}{208--213} (\bibinfo{year}{2006}).

\bibitem{katsaros2010hybrid}
\bibinfo{author}{Katsaros, G.} \emph{et~al.}
\newblock \bibinfo{title}{Hybrid superconductor--semiconductor devices made from self-assembled sige nanocrystals on silicon}.
\newblock \emph{\bibinfo{journal}{Nature nanotechnology}} \textbf{\bibinfo{volume}{5}}, \bibinfo{pages}{458--464} (\bibinfo{year}{2010}).

\bibitem{hendrickx2018}
\bibinfo{author}{Hendrickx, N.~W.} \emph{et~al.}
\newblock \bibinfo{title}{Gate-controlled quantum dots and superconductivity in planar germanium}.
\newblock \emph{\bibinfo{journal}{Nature Communications}} \textbf{\bibinfo{volume}{9}}, \bibinfo{pages}{2835} (\bibinfo{year}{2018}).

\bibitem{vigneau2019germanium}
\bibinfo{author}{Vigneau, F.} \emph{et~al.}
\newblock \bibinfo{title}{Germanium quantum-well josephson field-effect transistors and interferometers}.
\newblock \emph{\bibinfo{journal}{Nano letters}} \textbf{\bibinfo{volume}{19}}, \bibinfo{pages}{1023--1027} (\bibinfo{year}{2019}).

\bibitem{aggarwal2021enhancement}
\bibinfo{author}{Aggarwal, K.} \emph{et~al.}
\newblock \bibinfo{title}{Enhancement of proximity-induced superconductivity in a planar ge hole gas}.
\newblock \emph{\bibinfo{journal}{Physical Review Research}} \textbf{\bibinfo{volume}{3}}, \bibinfo{pages}{L022005} (\bibinfo{year}{2021}).

\bibitem{tosato2022hard}
\bibinfo{author}{Tosato, A.} \emph{et~al.}
\newblock \bibinfo{title}{Hard superconducting gap in germanium}.
\newblock \emph{\bibinfo{journal}{Communications Materials}} \textbf{\bibinfo{volume}{4}}, \bibinfo{pages}{23} (\bibinfo{year}{2023}).

\bibitem{Mayer2019}
\bibinfo{author}{Mayer, W.} \emph{et~al.}
\newblock \bibinfo{title}{Superconducting proximity effect in epitaxial al-inas heterostructures}.
\newblock \emph{\bibinfo{journal}{Applied Physics Letters}} \textbf{\bibinfo{volume}{114}}, \bibinfo{pages}{103104} (\bibinfo{year}{2019}).

\bibitem{likharev1979superconducting}
\bibinfo{author}{Likharev, K.}
\newblock \bibinfo{title}{Superconducting weak links}.
\newblock \emph{\bibinfo{journal}{Reviews of Modern Physics}} \textbf{\bibinfo{volume}{51}}, \bibinfo{pages}{101} (\bibinfo{year}{1979}).

\bibitem{rossner2003effective}
\bibinfo{author}{Rossner, B.}, \bibinfo{author}{Isella, G.} \& \bibinfo{author}{Kanel, H.~v.}
\newblock \bibinfo{title}{Effective mass in remotely doped ge quantum wells}.
\newblock \emph{\bibinfo{journal}{Applied Physics Letters}} \textbf{\bibinfo{volume}{82}}, \bibinfo{pages}{754--756} (\bibinfo{year}{2003}).

\bibitem{haxell2022large}
\bibinfo{author}{Haxell, D.~Z.} \emph{et~al.}
\newblock \bibinfo{title}{Measurements of phase dynamics in planar josephson junctions and squids}.
\newblock \emph{\bibinfo{journal}{Physical Review Letters}} \textbf{\bibinfo{volume}{130}}, \bibinfo{pages}{087002} (\bibinfo{year}{2023}).

\bibitem{reeg2018metallization}
\bibinfo{author}{Reeg, C.}, \bibinfo{author}{Loss, D.} \& \bibinfo{author}{Klinovaja, J.}
\newblock \bibinfo{title}{Metallization of a rashba wire by a superconducting layer in the strong-proximity regime}.
\newblock \emph{\bibinfo{journal}{Physical Review B}} \textbf{\bibinfo{volume}{97}}, \bibinfo{pages}{165425} (\bibinfo{year}{2018}).

\bibitem{fulton1972quantum}
\bibinfo{author}{Fulton, T.}, \bibinfo{author}{Dunkleberger, L.} \& \bibinfo{author}{Dynes, R.}
\newblock \bibinfo{title}{Quantum interference properties of double josephson junctions}.
\newblock \emph{\bibinfo{journal}{Physical Review B}} \textbf{\bibinfo{volume}{6}}, \bibinfo{pages}{855} (\bibinfo{year}{1972}).

\bibitem{paolucci2023gate}
\bibinfo{author}{Paolucci, F.}, \bibinfo{author}{De~Simoni, G.} \& \bibinfo{author}{Giazotto, F.}
\newblock \bibinfo{title}{A gate-and flux-controlled supercurrent diode effect}.
\newblock \emph{\bibinfo{journal}{Applied Physics Letters}} \textbf{\bibinfo{volume}{122}}, \bibinfo{pages}{042601} (\bibinfo{year}{2023}).

\bibitem{legg2023parity}
\bibinfo{author}{Legg, H.~F.}, \bibinfo{author}{Laubscher, K.}, \bibinfo{author}{Loss, D.} \& \bibinfo{author}{Klinovaja, J.}
\newblock \bibinfo{title}{Parity protected superconducting diode effect in topological josephson junctions}.
\newblock \emph{\bibinfo{journal}{arXiv preprint arXiv:2301.13740}}  (\bibinfo{year}{2023}).

\bibitem{lambert1997boundary}
\bibinfo{author}{Lambert, C.}, \bibinfo{author}{Raimondi, R.}, \bibinfo{author}{Sweeney, V.} \& \bibinfo{author}{Volkov, A.}
\newblock \bibinfo{title}{Boundary conditions for quasiclassical equations in the theory of superconductivity}.
\newblock \emph{\bibinfo{journal}{Physical Review B}} \textbf{\bibinfo{volume}{55}}, \bibinfo{pages}{6015} (\bibinfo{year}{1997}).

\bibitem{galaktionov2000second}
\bibinfo{author}{Galaktionov, A.} \& \bibinfo{author}{Ryu, C.-M.}
\newblock \bibinfo{title}{The second josephson harmonic in the dirty limit}.
\newblock \emph{\bibinfo{journal}{Journal of Physics: Condensed Matter}} \textbf{\bibinfo{volume}{12}}, \bibinfo{pages}{1351} (\bibinfo{year}{2000}).

\bibitem{ueda2020evidence}
\bibinfo{author}{Ueda, K.} \emph{et~al.}
\newblock \bibinfo{title}{Evidence of half-integer shapiro steps originated from nonsinusoidal current phase relation in a short ballistic inas nanowire josephson junction}.
\newblock \emph{\bibinfo{journal}{Physical Review Research}} \textbf{\bibinfo{volume}{2}}, \bibinfo{pages}{033435} (\bibinfo{year}{2020}).

\bibitem{zhang2022large}
\bibinfo{author}{Zhang, P.} \emph{et~al.}
\newblock \bibinfo{title}{Large second-order josephson effect in planar superconductor-semiconductor junctions}.
\newblock \emph{\bibinfo{journal}{arXiv preprint arXiv:2211.07119}}  (\bibinfo{year}{2022}).

\bibitem{iorio2023half}
\bibinfo{author}{Iorio, A.} \emph{et~al.}
\newblock \bibinfo{title}{Half-integer shapiro steps in highly transmissive {InSb} nanoflag josephson junctions}.
\newblock \emph{\bibinfo{journal}{Physical Review Research}} \textbf{\bibinfo{volume}{5}}, \bibinfo{pages}{033015} (\bibinfo{year}{2023}).
\newblock \bibinfo{note}{Publisher: American Physical Society}.

\bibitem{vanneste1988shapiro}
\bibinfo{author}{Vanneste, C.} \emph{et~al.}
\newblock \bibinfo{title}{Shapiro steps on current-voltage curves of dc squids}.
\newblock \emph{\bibinfo{journal}{Journal of applied physics}} \textbf{\bibinfo{volume}{64}}, \bibinfo{pages}{242--245} (\bibinfo{year}{1988}).

\bibitem{dartiailh2021missing}
\bibinfo{author}{Dartiailh, M.~C.} \emph{et~al.}
\newblock \bibinfo{title}{Missing shapiro steps in topologically trivial josephson junction on inas quantum well}.
\newblock \emph{\bibinfo{journal}{Nature communications}} \textbf{\bibinfo{volume}{12}}, \bibinfo{pages}{78} (\bibinfo{year}{2021}).

\bibitem{probst2015efficient}
\bibinfo{author}{Probst, S.}, \bibinfo{author}{Song, F.}, \bibinfo{author}{Bushev, P.~A.}, \bibinfo{author}{Ustinov, A.~V.} \& \bibinfo{author}{Weides, M.}
\newblock \bibinfo{title}{Efficient and robust analysis of complex scattering data under noise in microwave resonators}.
\newblock \emph{\bibinfo{journal}{Review of Scientific Instruments}} \textbf{\bibinfo{volume}{86}}, \bibinfo{pages}{024706} (\bibinfo{year}{2015}).

\bibitem{DegliEsposti2022}
\bibinfo{author}{Degli~Esposti, D.} \emph{et~al.}
\newblock \bibinfo{title}{Wafer-scale low-disorder 2deg in 28si/sige without an epitaxial si cap}.
\newblock \emph{\bibinfo{journal}{Applied Physics Letters}} \textbf{\bibinfo{volume}{120}}, \bibinfo{pages}{184003} (\bibinfo{year}{2022}).

\bibitem{Su}
\bibinfo{author}{Su, Y.-H.}, \bibinfo{author}{Chuang, Y.}, \bibinfo{author}{Liu, C.-Y.}, \bibinfo{author}{Li, J.-Y.} \& \bibinfo{author}{Lu, T.-M.}
\newblock \bibinfo{title}{Effects of surface tunneling of two-dimensional hole gases in undoped ge/gesi heterostructures}.
\newblock \emph{\bibinfo{journal}{Phys. Rev. Mater.}} \textbf{\bibinfo{volume}{1}}, \bibinfo{pages}{044601} (\bibinfo{year}{2017}).

\bibitem{stehouwer2023germanium}
\bibinfo{author}{Stehouwer, L. E.~A.} \emph{et~al.}
\newblock \bibinfo{title}{Germanium wafers for strained quantum wells with low disorder}.
\newblock \emph{\bibinfo{journal}{Applied Physics Letters}} \textbf{\bibinfo{volume}{123}}, \bibinfo{pages}{092101} (\bibinfo{year}{2023}).

\bibitem{Leijnse2013}
\bibinfo{author}{Leijnse, M.} \& \bibinfo{author}{Flensberg, K.}
\newblock \bibinfo{title}{Coupling spin qubits via superconductors}.
\newblock \emph{\bibinfo{journal}{Phys. Rev. Lett.}} \textbf{\bibinfo{volume}{111}}, \bibinfo{pages}{060501} (\bibinfo{year}{2013}).

\bibitem{spethmann2023}
\bibinfo{author}{Spethmann, M.}, \bibinfo{author}{Bosco, S.}, \bibinfo{author}{Hofmann, A.}, \bibinfo{author}{Klinovaja, J.} \& \bibinfo{author}{Loss, D.}
\newblock \bibinfo{title}{High-fidelity two-qubit gates of hybrid superconducting-semiconducting singlet-triplet qubits}.
\newblock \emph{\bibinfo{journal}{arXiv preprint arXiv:2304.05086}}  (\bibinfo{year}{2023}).

\bibitem{Burkard2020}
\bibinfo{author}{Burkard, G.}, \bibinfo{author}{Gullans, M.~J.}, \bibinfo{author}{Mi, X.} \& \bibinfo{author}{Petta, J.~R.}
\newblock \bibinfo{title}{Superconductor--semiconductor hybrid-circuit quantum electrodynamics}.
\newblock \emph{\bibinfo{journal}{Nature Reviews Physics}} \textbf{\bibinfo{volume}{2}}, \bibinfo{pages}{129--140} (\bibinfo{year}{2020}).

\bibitem{ciaccia2023charge}
\bibinfo{author}{Ciaccia, C.} \emph{et~al.}
\newblock \bibinfo{title}{Charge-4e supercurrent in an inas-al superconductor-semiconductor heterostructure}.
\newblock \emph{\bibinfo{journal}{arXiv preprint arXiv:2306.05467}}  (\bibinfo{year}{2023}).

\bibitem{li2023interfering}
\bibinfo{author}{Li, Y.} \emph{et~al.}
\newblock \bibinfo{title}{Interfering josephson diode effect and magnetochiral anisotropy in ta2pd3te5 asymmetric edge interferometer}.
\newblock \emph{\bibinfo{journal}{arXiv preprint arXiv:2306.08478}}  (\bibinfo{year}{2023}).

\bibitem{matsuo2023josephson}
\bibinfo{author}{Matsuo, S.} \emph{et~al.}
\newblock \bibinfo{title}{Josephson diode effect derived from short-range coherent coupling}.
\newblock \emph{\bibinfo{journal}{Nature Physics}} \textbf{\bibinfo{volume}{19}}, \bibinfo{pages}{1636--1641} (\bibinfo{year}{2023}).
\newblock \bibinfo{note}{Number: 11 Publisher: Nature Publishing Group}.

\bibitem{jirovec2021singlet}
\bibinfo{author}{Jirovec, D.} \emph{et~al.}
\newblock \bibinfo{title}{A singlet-triplet hole spin qubit in planar ge}.
\newblock \emph{\bibinfo{journal}{Nature materials}} \textbf{\bibinfo{volume}{20}}, \bibinfo{pages}{1106--1112} (\bibinfo{year}{2021}).

\bibitem{annunziata2010tunable}
\bibinfo{author}{Annunziata, A.~J.} \emph{et~al.}
\newblock \bibinfo{title}{Tunable superconducting nanoinductors}.
\newblock \emph{\bibinfo{journal}{Nanotechnology}} \textbf{\bibinfo{volume}{21}}, \bibinfo{pages}{445202} (\bibinfo{year}{2010}).

\bibitem{courtois2008origin}
\bibinfo{author}{Courtois, H.}, \bibinfo{author}{Meschke, M.}, \bibinfo{author}{Peltonen, J.} \& \bibinfo{author}{Pekola, J.~P.}
\newblock \bibinfo{title}{Origin of hysteresis in a proximity josephson junction}.
\newblock \emph{\bibinfo{journal}{Physical review letters}} \textbf{\bibinfo{volume}{101}}, \bibinfo{pages}{067002} (\bibinfo{year}{2008}).

\end{thebibliography}

\clearpage

\section*{Extended data}
\renewcommand{\thefigure}{ED\arabic{figure}}
\setcounter{figure}{0}
\renewcommand{\thetable}{ED\arabic{table}}
\setcounter{table}{0}

\begin{figure} [h!]
\centering
  \includegraphics[]{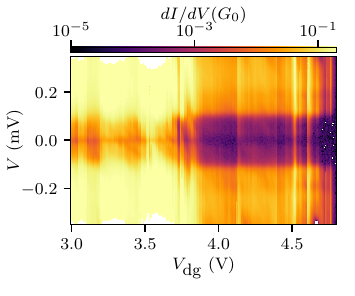}
  \caption{\textbf{Additional data for Fig.2.} \small{  $dI/dV$ as a function of $V$ and $V_{\textrm{dg}}$ plotted in logarithmic scale for sample D5. The data for less positive $V_{\textrm{dg}}$ show indications of quantum dots formation.} } \label{figure2_SI}
\end{figure}

\begin{table*}[h]
\centering
\begin{tabular}{|*{7}{l|}}
\hline
sample & $f_{\textrm{r}} ($\unit{GHz}$)$ & $Q_{\textrm{e}}$ & $Q_{\textrm{i},\langle n_{\textrm{ph}}\rangle = 0.1}$ & $Q_{\textrm{i},\langle n_{\textrm{ph}}\rangle = 1}$ & $Q_{\textrm{i},\langle n_{\textrm{ph}}\rangle = 10}$ & $Q_{\textrm{i},\langle n_{\textrm{ph}}\rangle = 100}$ \\ [0.5ex] 
\hline
without QW & $4.01$ & $6150 \pm 150$ & $2800 \pm 200$ & $6000 \pm 300$ & $13000 \pm 600$ & $30000 \pm 3000$ \\ [0.5ex]
\hline
without QW & $4.04$ & $6300 \pm 150$ & n.a. & $6270 \pm 130$ & $10000 \pm 130$ & $15700 \pm 170$ \\ [0.5ex] 
\hline
without QW & $4.18$ & $3250 \pm 70$ & n.a. & $9240 \pm 350$ & $15350 \pm 450$ & $26350 \pm 635$ \\ [0.5ex] 
\hline
without QW & $4.23$ & $3380 \pm 40$ & n.a. & $5670 \pm 140$ & $8220 \pm 125$ & $11000 \pm 175$ \\ [0.5ex] 
\hline
without QW & $4.27$ & $3800 \pm 150$ & $2700 \pm 200$ & $7000 \pm 300$ & $15000 \pm 800$ & $26000 \pm 3000$ \\ [0.5ex] 
\hline
without QW & $4.60$ & $2750 \pm 150$ & $3800 \pm 200$ & $8800 \pm 300$ & $17000 \pm 500$ & $24000 \pm 1000$ \\ [0.5ex] 
\hline
without QW & $4.66$ & $4000 \pm 300$ & $5200 \pm 500$ & $10000 \pm 500$ & $21000 \pm 700$ & $40000 \pm 2000$ \\ [0.5ex] 
\hline
without QW & $4.77$ & $2000 \pm 40$ & n.a. & $7780 \pm 200$ & $11900 \pm 400$ & $17100 \pm 1120$ \\ [0.5ex] 
\hline
without QW & $4.84$ & $1370 \pm 15$ & n.a. & $7600 \pm 360$ & $12220 \pm 1220$ & $18350 \pm 3000$ \\ [0.5ex] 
\hline
without QW & $4.91$ & $2500 \pm 200$ & $5000 \pm 400$ & $10000 \pm 600$ & $17000 \pm 1000$ & $27000 \pm 2000$ \\ [0.5ex] 
\hline
without QW & $5.02$ & $2100 \pm 70$ & $3100 \pm 100$ & $4300 \pm 100$ & $6560 \pm 240$ & $9000 \pm 725$ \\ [0.5ex]
\hline
without QW & $5.24$ & $1550 \pm 20$ & $4100 \pm 175$ & $5340 \pm 160$ & $7430 \pm 300$ & $9640 \pm 600$ \\ [0.5ex]
\hline
without QW & $5.72$ & $1660 \pm 25$ & $4300 \pm 135$ & $5500 \pm 145$ & $7660 \pm 260$ & $10200 \pm 930$ \\ [0.5ex]
\hline
without QW & $5.78$ & $1365 \pm 20$ & $4830 \pm 180$ & $5950 \pm 140$ & $8230 \pm 220$ & $10270 \pm 500$ \\ [0.5ex] 
\hline
without QW & $6.00$ & $1700 \pm 20$ & $5250 \pm 220$ & $6300 \pm 150$ & $9060 \pm 230$ & $12580 \pm 500$ \\ [0.5ex]
\hline
D5 & $4.04$ & $4740 \pm 800$ & n.a. & $5365 \pm 225$ & $8920 \pm 550$ & $14360 \pm 1240$ \\ [0.5ex]
\hline
D5 & $4.13$ & $3200 \pm 215$ & n.a. & $6400 \pm 260$ & $10530 \pm 825$ & $15400 \pm 1000$ \\ [0.5ex]
\hline
D5 & $4.22$ & $3300 \pm 130$ & n.a. & $3230 \pm 230$ & $11000 \pm 610$ & $15750 \pm 1160$ \\ [0.5ex]
\hline
D5 & $4.35$ & $3085 \pm 215$ & n.a. & $5850 \pm 230$ & $10150 \pm 570$ & $15200 \pm 1200$ \\ [0.5ex]
\hline
D5 & $4.39$ & $1440 \pm 55$ & n.a. & $5690 \pm 220$ & $10400 \pm 680$ & $14200 \pm 1750$ \\ [0.5ex]
\hline
D5 & $4.43$ & $2100 \pm 100$ & n.a. & $5500 \pm 210$ & $8900 \pm 470$ & $13040 \pm 720$ \\ [0.5ex] 
\hline
D5 & $4.52$ & $1680 \pm 240$ & n.a. & $5230 \pm 510$ & $6880 \pm 1060$ & $10000 \pm 3050$ \\ [0.5ex] 
\hline
D5 & $4.56$ & $2450 \pm 170$ & n.a. & $7100 \pm 220$ & $12600 \pm 1530$ & $18150 \pm 3700$ \\ [0.5ex] 
\hline
D5 & $4.61$ & $1250 \pm 50$ & n.a. & $6500 \pm 400$ & $20000 \pm 4000$ & $36000 \pm 17000$ \\ [0.5ex] 
\hline
D5 & $4.73$ & $800 \pm 20$ & $4000 \pm 200$ & $7600 \pm 500$ & $14000 \pm 2000$ & $23000 \pm 5000$ \\ [0.5ex]
\hline
D5 & $5.23$ & $770 \pm 25$ & $2730 \pm 200$ & $4545 \pm 340$ & $6980 \pm 1230$ & $8530 \pm 1400$ \\ [0.5ex]
\hline
D5 & $5.45$ & $740 \pm 25$ & $3600 \pm 200$ & $5500 \pm 200$ & $10000 \pm 900$ & $17000 \pm 1000$ \\ [0.5ex]
\hline
D5 & $5.83$ & $970 \pm 20$ & $1645 \pm 70$ & $2160 \pm 70$ & $2680 \pm 100$ & $2910 \pm 170$ \\ [0.5ex] 
\hline
D5 & $5.97$ & $630 \pm 10$ & n.a. & $2900 \pm 150$ & $3400 \pm 100$ & $4000 \pm 200$ \\ [0.5ex] 
\hline
D60 & $4.47$ & $3220 \pm 50$ & n.a. & $4460 \pm 75$ & $5660 \pm 50$ & $6600 \pm 110$ \\ [0.5ex]
\hline
D60 & $4.70$ & $2870 \pm 50$ & n.a. & $3365 \pm 65$ & $4130 \pm 90$ & $4830 \pm 110$ \\ [0.5ex]
\hline
D60 & $4.98$ & $2745 \pm 145$ & $2700 \pm 75$ & $3200 \pm 35$ & $3700 \pm 30$ & $4100 \pm 65$ \\ [0.5ex]
\hline
D60 & $5.09$ & $2610 \pm 160$ & $2910 \pm 130$ & $3175 \pm 55$ & $3570 \pm 65$ & $4000 \pm 130$ \\ [0.5ex] 
\hline
D60 & $5.24$ & $2100 \pm 30$ & $2440 \pm 60$ & $2870 \pm 45$ & $3260 \pm 50$ & $3470 \pm 60$ \\ [0.5ex] 
\hline
D60 & $5.34$ & $683 \pm 5$ & $2300 \pm 80$ & $2555 \pm 35$ & $2840 \pm 50$ & $3070 \pm 60$ \\ [0.5ex] 
\hline
D60 & $5.68$ & $627 \pm 3$ & $2040 \pm 70$ & $2240 \pm 30$ & $2530 \pm 20$ & $2670 \pm 20$ \\ [0.5ex] 
\hline
D60 & $6.14$ & $730 \pm 10$ & $1740 \pm 60$ & $1930 \pm 40$ & $2100 \pm 40$ & $2200 \pm 50$ \\ [0.5ex] 
\hline
\end{tabular}
\caption{\textbf{Summary of the resonator data.} \small{ Resonance frequency $f_{\textrm{r}}$, external quality factor $Q_{\textrm{e}}$ and internal quality factor $Q_{\textrm{i}}$ for different average photon number $\langle n_{\textrm{ph}} \rangle$ extracted using an algebraic fit of the parameter $S_{21}$ \cite{probst2015efficient}. The values reported here are average values and the error is given by the standard deviation. The error of $f_{\textrm{r}}$ is not reported because it is smaller than $\qty{1}{MHz}$. n.a. stands for not available. We note that higher frequency resonators show lower $Q_{\textrm{i}}$. } }
\label{tab:resonator}
\end{table*}

\begin{figure*} 
  \includegraphics[]{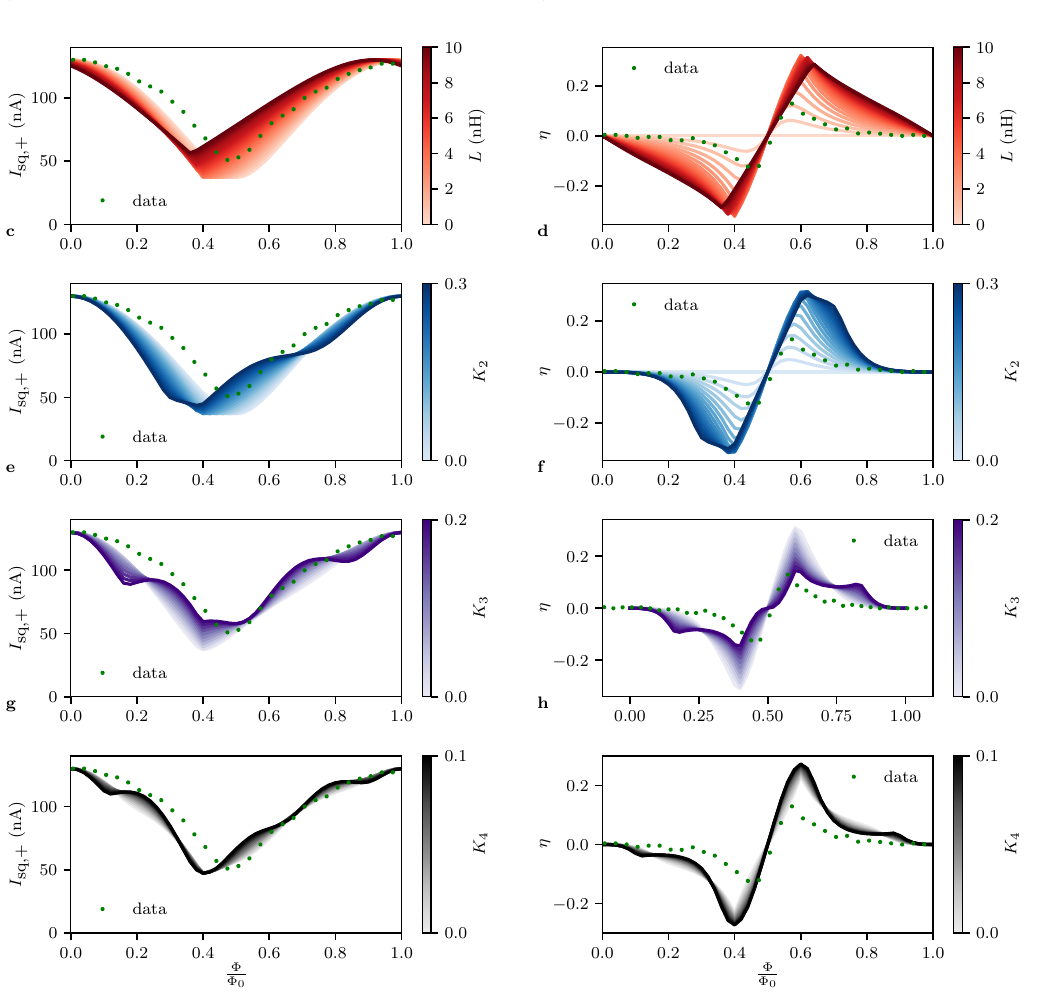}
  \caption{\textbf{SQUID behaviour for different cases.} \small{ Left column [right column] $I_{\textrm{sq,+}}$  [$\eta$] as a function of $\Phi$. The green dots represent the experimental values extracted from Fig.~\ref{figure4}\textbf{b}. \textbf{a} [\textbf{b}] represent the theoretical calculation for different values of $L$ and ignoring higher harmonic contribution. The blue traces in \textbf{c-d} represent the theoretical calculation using $L = 0$, $K_1=1-K_2, K_3=K_4=0$ and $K_2$ varies from $0$ to $0.3$. The purple traces in \textbf{e-f} are for $L = 0$, $K_1=0.83-K_3, K_2=0.17, K_4=0$ and $K_3$ varies from $0$ to $0.2$. The black traces in \textbf{g-h} are for $L = 0$, $K_1=0.73-K_4, K_2=0.17, K_3=0.1$ and $K_4$ varies from $0$ to $0.1$.} }\label{FigureSI_L_harm_dependence}
\end{figure*}



\begin{figure*} [h!]
  \includegraphics[]{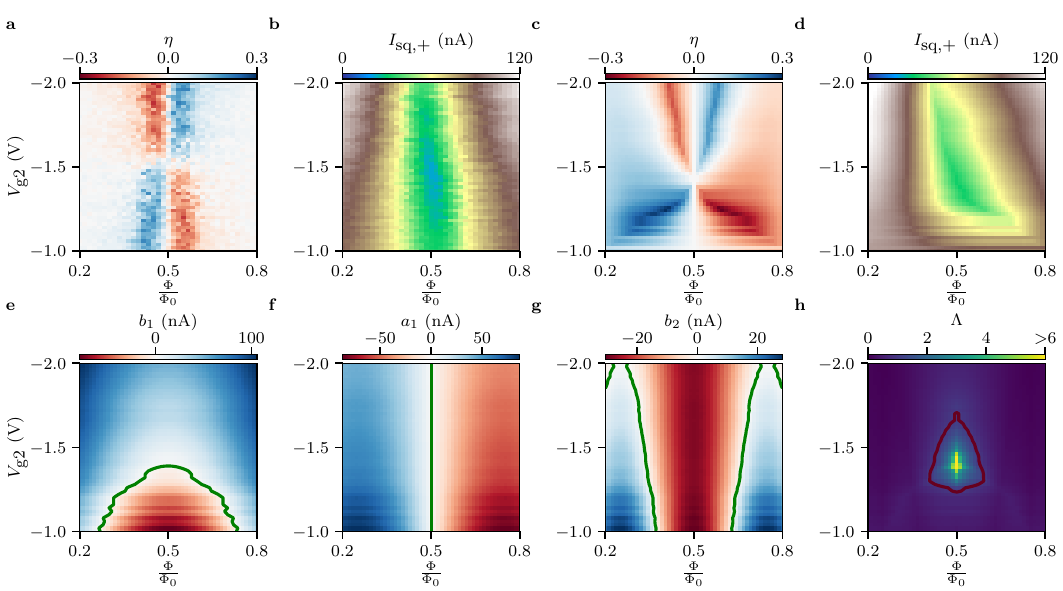}
  \caption{\textbf{Expected results with asymmetric harmonic contributions.} \small{ \textbf{a} [\textbf{b}] shows the same plot as Fig.~\ref{figure4}\textbf{e} [Fig.~\ref{figure4}\textbf{f}]. \textbf{c} [\textbf{d}] Theoretical prediction of \textbf{a} [\textbf{b}], where, like in the main text, it was assumed that $L = \qty{110}{pH}$  and $V_{\textrm{g1}} = \qty{-1.5}{V}$. The harmonic contributions of $I_{\textrm{JJ2}}$ are assumed to be $K_1=0.7,K_2=0.1,K_3=0.1$ and $I_{\textrm{JJ1}}$  was assumed to have more higher harmonic contribution, namely $I_{\textrm{JJ1}}$ has $K_1=0.6,K_2=0.2,K_3=0.1$. \textbf{e-f} First-order  harmonics contribution as a function of  $V_{\textrm{g2}}$ and $\Phi$.  Both terms vanish close to the balanced point and at $\frac{\Phi}{\Phi_0} = 0.5$. \textbf{g} Sinusoidal second harmonic contribution as a function of $V_{\textrm{g2}}$  and $\Phi$. Importantly, $b_2$ never vanishes at $\frac{\Phi}{\Phi_0} = 0.5$. \textbf{h} Ratio between second and first harmonic as a function of  $V_{\textrm{g2}}$ and $\Phi$. For this situation, the second component would dominate the CPR not only at the sweet spot. Therefore a small asymmetry between the harmonic contributions would not change the fact that the second harmonic contribution never vanishes at $\frac{\Phi}{\Phi_0} = 0.5$.  In \textbf{c-h} the same parameters have been used.}}  \label{figureSI_asymmetrysmall}
\end{figure*}

\begin{figure*} [h!]
  \includegraphics[]{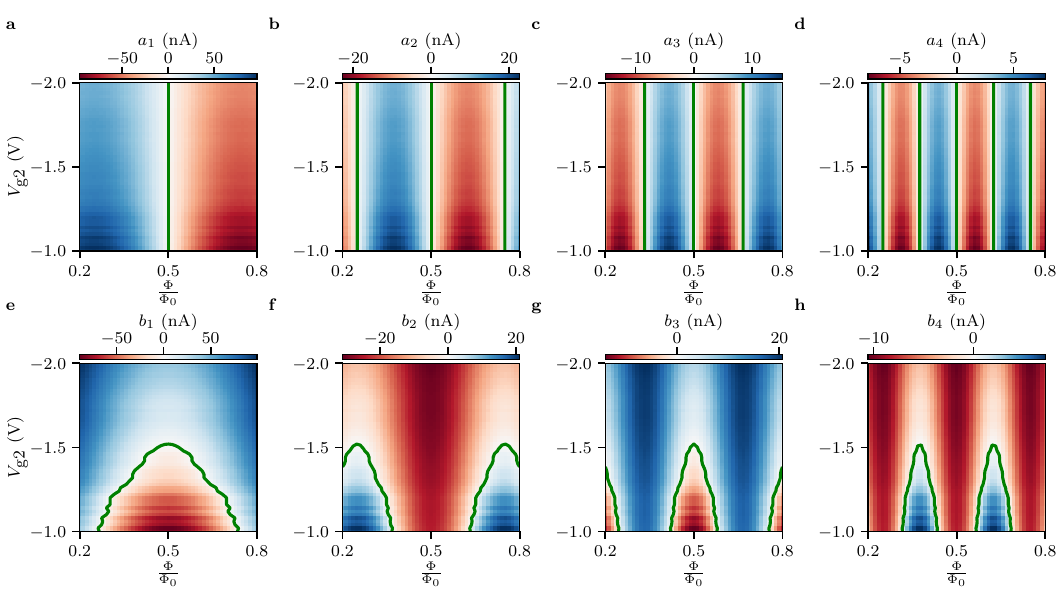}
  \caption{\textbf{Higher harmonic contribution as a function of $\Phi$ and $V_{\textrm{g2}}$ estimated using the same parameters as in Fig.~\ref{figure4}.} \small{ Upper [lower] row shows the cosinus [sinus] terms. Interestingly, at the sweet spot ($V_{\textrm{g1}} = V_{\textrm{g2}}$ and $\frac{\Phi}{\Phi_0} =0.5$)} all terms, but $b_2$ and $b_4$, vanish. }\label{figure_SI_harmonics}
\end{figure*}

\begin{figure*} [h!]
  \includegraphics[]{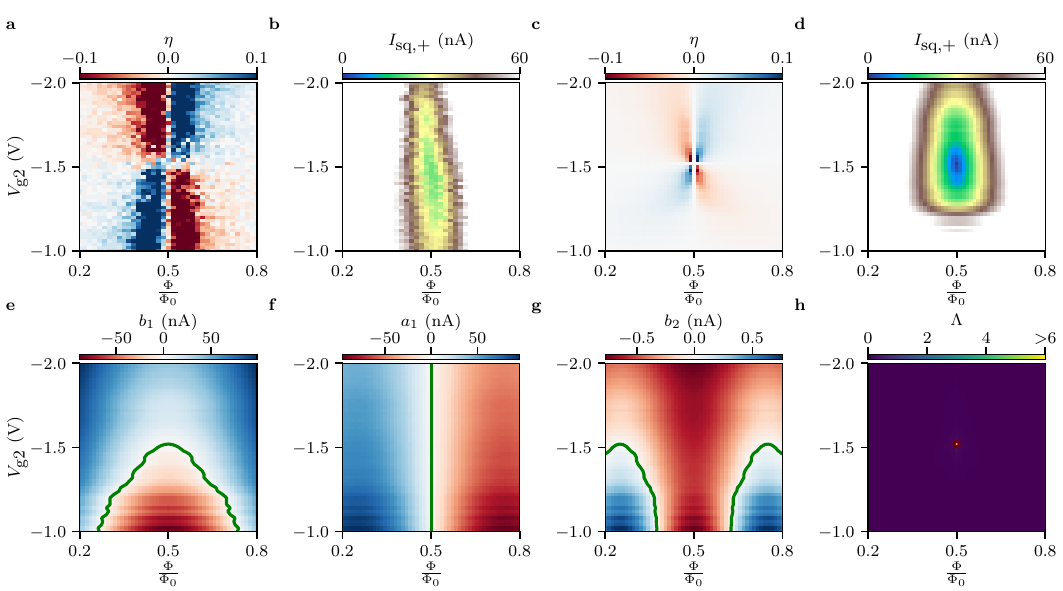}
  \caption{\textbf{Expected results with standard sinusoidal CPRs and low inductance ($\boldsymbol{ L = \qty{110}{pH}}$)} \small{ \textbf{a} [\textbf{b}] shows \ the same plot as Fig.~\ref{figure4}\textbf{e} [Fig.~\ref{figure4}\textbf{f}]. \textbf{c} [\textbf{d}] Theoretical calculation of \textbf{a} [\textbf{b}], assuming standard sinusoidal CPR and $\qty{110}{pH}$. The theoretical prediction does not resemble the experimental data. In \textbf{c}, $\left| \eta \right|$ is much smaller than what has been experimentally observed (\textbf{a}) and $\eta$ is finite only slightly away from sweet spot, located at $V_{\textrm{g2}} = \qty{-1.5}{V}$ and at $\Phi = 0.5 \Phi_0$. In addition, the expected $I_{\textrm{sq,+}}$ is much lower than the measured values, especially at the sweet spot. In the theoretical calculation, $I_{\textrm{sq,+}} \approx 0$ at the sweet spot (\textbf{d}); whereas in the experimental data $I_{\textrm{sq,+}} \approx \qty{20}{nA}$ (\textbf{b}). \textbf{e-f} First-order harmonics contribution as a function of $I_{\textrm{ret2}}$ and $\Phi$  extracted using the same parameters as in \textbf{c-d}. Like for the case described in the main text, at the balanced point and at $\frac{\Phi}{\Phi_0} = 0.5$ both first harmonic terms vanish. \textbf{g} Second sinusoidal harmonics contribution as a function of $I_{\textrm{ret2}}$ and $\Phi$. Importantly, $b_2$ never vanishes at $\frac{\Phi}{\Phi_0} = 0.5$. Despite the fact that it is qualitatively similar to the situation of the main text (Fig.~\ref{figure4}\textbf{k}), there is a major difference. At the sweet spot $\left|b_2\right| \approx 0.1 \ \si{nA}$, while in Fig.~\ref{figure4}\textbf{k} $\left|b_2\right| \approx 20 \ \si{nA}$. \textbf{h} Ratio between second and first harmonic. For this situation, the second component would dominate the CPR just at the sweet spot.} } \label{FigureSI_Lsmall}
\end{figure*}

\begin{figure*} [h!]
  \includegraphics[]{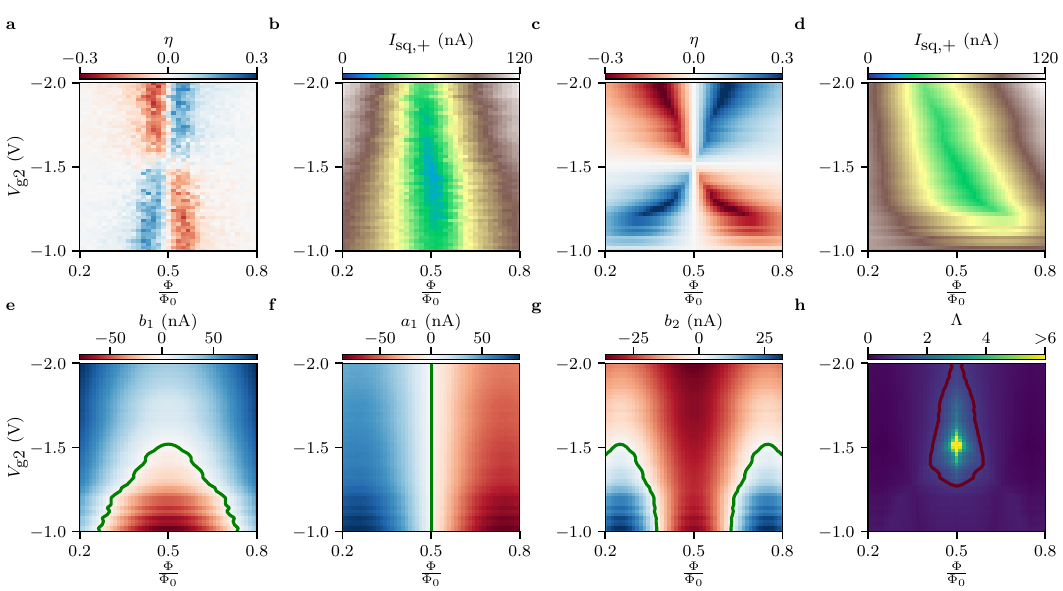}
  \caption{\textbf{Expected results with standard sinusoidal CPRs and with an inductance ($\boldsymbol{ L = \qty{6}{nH}}$) more than one order of magnitude higher than what was measured.} \small{ \textbf{a} [\textbf{b}] shows \ the same plot as Fig.~\ref{figure4}\textbf{e} [Fig.~\ref{figure4}\textbf{f}]. \textbf{c} [\textbf{d}] Theoretical calculation of \textbf{a} [\textbf{b}], assuming standard sinusoidal CPRs and $L = \qty{6}{nH}$, which is much higher than what has been experimentally estimated. The theoretical calculation resembles qualitatively the experimental data. \textbf{e-f} First-order harmonics contribution as a function of $V_{\textrm{g2}}$ and $\Phi$  extracted using the same parameters as in \textbf{c-d}. Similar to the case described in the main text, close to the balanced point and at $\frac{\Phi}{\Phi_0} = 0.5$ both first harmonic terms vanish. \textbf{g} Second sinusoidal harmonics contribution as a function of $V_{\textrm{g2}}$ and $\Phi$. Importantly, $b_2$ never vanishes at half flux quantum.  \textbf{h} Ratio between second and first harmonic. For this situation, the second component would dominate the CPR not only at the sweet spot, but also in the area enclosed by the red line corresponding to $K=1$. } }\label{FigureSI_Lbig}
\end{figure*}

\begin{figure*} [h!]
\centering
  \includegraphics[]{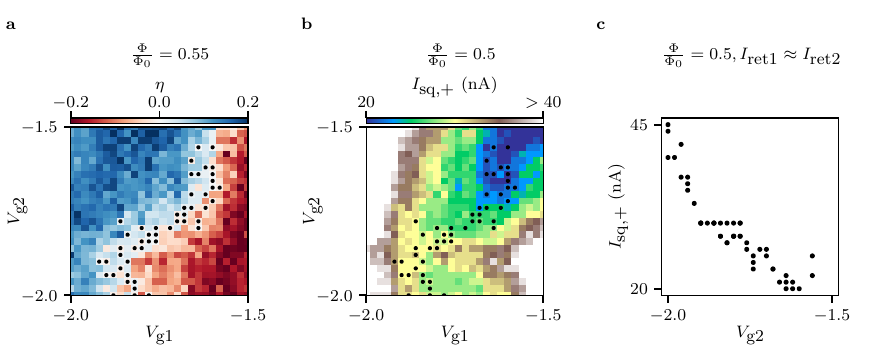}
  \caption{\textbf{Procedure to identify and quantify the region of the sweet spot.} \small{ \textbf{a} $\eta$  for $\frac{\Phi}{\Phi_0} = 0.55$ as a function of $V_{\textrm{g1}}$ and $V_{\textrm{g2}}$. If $\left| V_{\textrm{g2}} \right| > \left| V_{\textrm{g1}} \right|$ [$\left| V_{\textrm{g2}} \right| < \left| V_{\textrm{g1}} \right|$] $\eta < 0$ [$\eta > 0$]. Importantly, $\eta$ vanishes along the diagonal line, i.e. if $V_{\textrm{g1}} \approx V_{\textrm{g2}}$ which corresponds to $I_{\textrm{ret1}} \approx I_{\textrm{ret2}}$. The black dots indicate the points at which $\left| \eta \right|  < 0.015$. Once the $I_{\textrm{ret1}} \approx I_{\textrm{ret2}}$ region is identified, we move to $\frac{\Phi}{\Phi_0} = 0.5$ in order to be able to assess the sweet spot regime. \textbf{b} shows $I_{\textrm{sq,+}}$ as a function of $V_{\textrm{g1}}$ and $V_{\textrm{g2}}$. $I_{\textrm{sq,+}}$ is finite even when the junctions are balanced, see black spots. \textbf{c} $I_{\textrm{sq,+}}$ as a function of $V_{\textrm{g2}}$ in the sweet spot configuration. The identification of the balanced region cannot be carried out at half flux quantum because $\eta$ is always zero.} }  \label{FigureSI_gatevsgate}
\end{figure*}

\begin{figure} [h!]
  \includegraphics[]{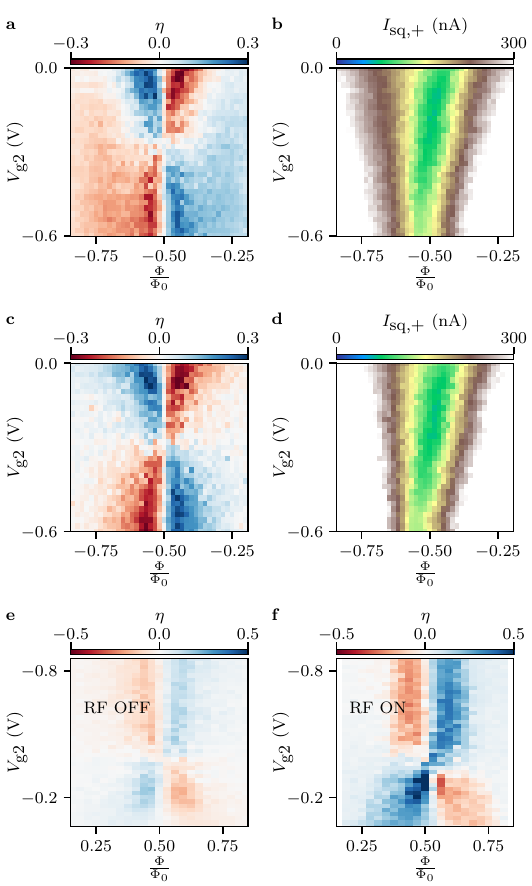}
  \caption{\textbf{Additional SDE results.} \small{ \textbf{a} [\textbf{b}] $\eta$ [$I_{\textrm{sq,+}}$]  as a function of $\Phi$ and $V_{\textrm{g2}}$ for a $\qty{30}{nm}$-thick Al device.  In \textbf{a} $\eta$ is always zero at $ \frac{\Phi}{\Phi_0} = 0.5$, independently on $V_{\textrm{g2}}$, and the polarity of the diode is inverted at the balanced point. Differently from Fig.~\ref{figure4}\textbf{e}, the SDE does not vanish for all values of $\Phi$ at the balanced point. This can be due to an asymmetric harmonic contribution (see Fig.~\ref{figureSI_asymmetrysmall}\textbf{c}).  In \textbf{a} [\textbf{b}],  $\eta$ [$I_{\textrm{sq,+}}$] was extracted by recording the retrapping current for both branches. \textbf{c} [\textbf{d}] is the same measurement as \textbf{a} [\textbf{b}] but now the switching current was recorded.  Importantly,  both approaches give similar results. \textbf{e} $ \eta$ as a function of $\Phi$ and $V_{\textrm{g2}}$ for another sample D8 with thin film of Al deposited at low temperature. \textbf{f} same as \textbf{e} but with the addition of an applied RF power, namely with $P = \qty{-40}{dBm}$ and $f_{\textrm{ac}} = \qty{1.5}{GHz}$. Importantly $\eta$ increases with an applied drive. } } \label{FigureSI_more_diode}
\end{figure}

\begin{figure*}
    \centering   \includegraphics{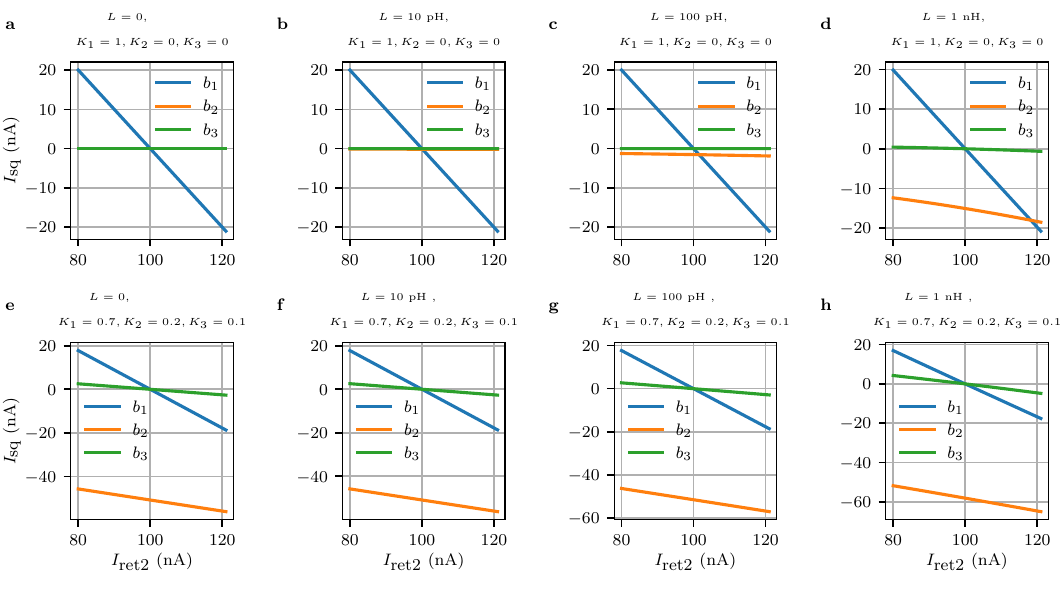}
   \caption{\textbf{Calculation of the first three harmonics of the SQUID CPR at half flux quantum by varying $I_{\textrm{ret2}}$ and with $I_{\textrm{ret1}} = 100$ nA.} In the upper row, higher order term are neglected ($K_2=K_3=0$). In the lower row, higher order terms are considered ($K_1=0.7, K_2=0.2, K_3 =0.1$). In the first column it is assumed $L=0$, in the second $L=10 $ pH (as for the thick-Aluminum samples), in the third  $L=100 $ pH (as for the thin-Aluminum samples) and in the fourth row $L=1 $ nH.}
    \label{fig:reply2}
\end{figure*}

\begin{figure*} [h!]
  \includegraphics[]{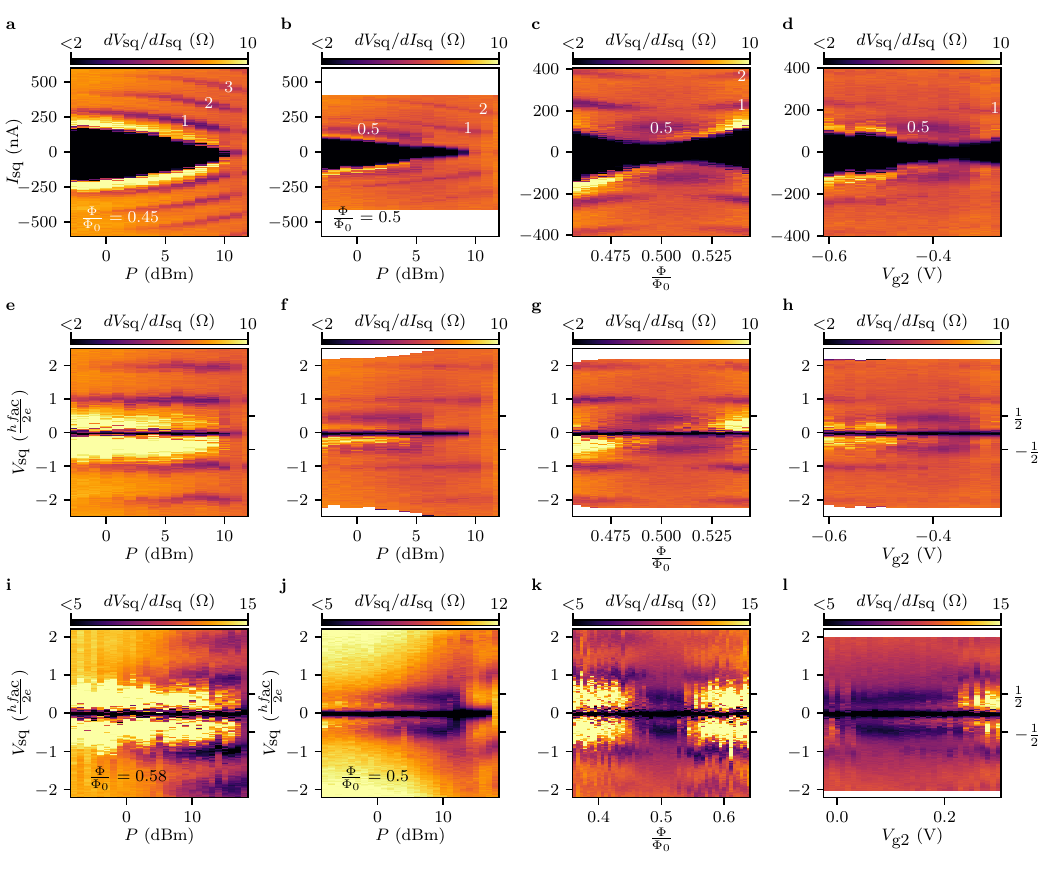}
  \caption{\textbf{Observation of half-integer steps for another device and identification of Shapiro steps.} \small{ \textbf{a} [\textbf{b}] Shapiro pattern for a $\qty{30}{nm}$ thick room temperature deposited Al sample  with $R_{\textrm{shunt}} = \qty{10}{\Omega}$ in the balanced regime ($I_{\textrm{ret1}} \approx I_{\textrm{ret2}} \approx \qty{400}{nA}$)   with $f_{\textrm{ac}} = \qty{500}{MHz}$ and at $\frac{\Phi}{\Phi_0} = 0.45$ [$\frac{\Phi}{\Phi_0} = 0.5$]. The differential resistance $dV_{\textrm{sq}}/dI_{\textrm{sq}}$ is plotted as a function of the RF power $P$ and $I_{\textrm{sq}}$. Dips in $dV_{\textrm{sq}}/dI_{\textrm{sq}}$ correspond to integer Shapiro steps. Importantly at half flux quantum (\textbf{b}), the first half-integer step appears for low $P$, see white numbers.  \textbf{c} Shapiro map as a function of $I_{\textrm{sq}}$ and $\Phi$ in the balanced regime (like for the previous plots) for $P = \qty{4}{dBm}$. The half-integer steps appear only close to $\frac{\Phi}{\Phi_0}=0.5$. \textbf{d} Shapiro map as a function of $I_{\textrm{sq}}$ and $V_{\textrm{g2}}$ for $P = \qty{4}{dBm}$  and at $\frac{\Phi}{\Phi_0}=0.5$. The half-integer step appears when the SQUID is close to the balanced condition, i.e. if $I_{\textrm{ret2}} \approx 400 \pm 25 \unit{nA}$. The white numbers in \textbf{c} and in \textbf{d} emphasize the appearance of the half-integer step. The second row corresponds to the same measurements as the first row but they are plotted as a function of $V_{\textrm{sq}}$ instead of $I_{\textrm{sq}}$. The third row corresponds to the device shown in the main text in Fig.~\ref{figure5} but the data are plotted as a function of $V_{\textrm{sq}}$ as well. \textbf{i} corresponds to Fig.~\ref{figure5}\textbf{b}, \textbf{j} corresponds to Fig.~\ref{figure5}\textbf{d}, \textbf{k} corresponds to Fig.~\ref{figure5}\textbf{e} and \textbf{l} corresponds to Fig.~\ref{figure5}\textbf{f}. $V_{\textrm{sq}}$ is obtained by integrating the signal of $dV_{\textrm{sq}}/dI_{\textrm{sq}}$. } } \label{figureSI_shapiro_10Ohm}
\end{figure*}

\end{document}